\documentclass[reprint,superscriptaddress, prb,
 amsmath,amssymb,aps,floatfix]{revtex4-2}
\usepackage{graphicx}
\usepackage{dcolumn}
\usepackage{bm}
\usepackage{hyperref}

\usepackage{xcolor}
\usepackage[normalem]{ulem}
\usepackage{xr}
\usepackage{booktabs}
\usepackage{multirow}
\usepackage{amsmath}
\usepackage{braket}
\usepackage{amssymb}
\usepackage{float}

\makeatletter
\newcommand*{\addFileDependency}[1]{
  \typeout{(#1)}
  \@addtofilelist{#1}
  \IfFileExists{#1}{}{\typeout{No file #1.}}
}
\makeatother

\newcommand{\tmop}[1]{\ensuremath{\operatorname{#1}}}

\newcommand{\matrixf}[1]{\underline{\mathbf{#1}}}

\newcommand{\tmpc}{\matrixf{T}_{PC}}

\newcommand{\tmnao}{\matrixf{\tilde{T}}}

\newcommand{\aoeq}[2]{| \varphi_{#1 #2} \rangle }
\newcommand{\blochaoeq}[2]{| \chi_{\tmop{#1 #2}} \rangle }

\renewcommand{\vec}[1]{\mathbf{#1}}

\begin{document}

\preprint{APS/123-QED}

\title{Efficient Band Structure Unfolding with Atom-centered Orbitals:\\ General Theory and Application}

\author{Jingkai Quan}
\affiliation{
 The NOMAD Laboratory at Fritz-Haber-Institut der Max-Planck-Gesellschaft, Faradayweg 4-6, D-14195, Berlin, Germany
}
\affiliation{
 Max-Planck Institute for the Structure and Dynamics of Matter, Luruper Chausse 149, 22761, Hamburg, Germany
 }
\author{Nikita~Rybin}
\affiliation{
 The NOMAD Laboratory at Fritz-Haber-Institut der Max-Planck-Gesellschaft, Faradayweg 4-6, D-14195, Berlin, Germany
}

\author{Matthias Scheffler}
\affiliation{
 The NOMAD Laboratory at Fritz-Haber-Institut der Max-Planck-Gesellschaft, Faradayweg 4-6, D-14195, Berlin, Germany
}
\author{Christian Carbogno}
\email{carbogno@fhi-berlin.mpg.de}
\affiliation{
 The NOMAD Laboratory at Fritz-Haber-Institut der Max-Planck-Gesellschaft, Faradayweg 4-6, D-14195, Berlin, Germany
}

\date{\today}

\begin{abstract}
Band structure unfolding is a key technique for analyzing and simplifying the electronic band structure of large, internally distorted supercells that break the primitive cell's translational symmetry. In this work, we present an efficient band unfolding method for atomic orbital (AO) basis sets that explicitly accounts for both the non-orthogonality of atomic orbitals and their atom-centered nature. Unlike existing approaches that typically rely on a plane-wave representation of the (semi-)valence states, we here derive analytical expressions that recasts the primitive cell translational operator and the associated Bloch-functions in the supercell AO basis.  In turn, this enables the accurate and efficient unfolding of conduction, valence, and core states in all-electron codes, as demonstrated by our implementation in the all-electron {\it ab initio} simulation package {\tt FHI-aims}, which employs numeric atom-centered orbitals. We explicitly demonstrate the capability of running large-scale unfolding calculations for systems with thousands of atoms and showcase the importance of this technique for computing temperature-dependent spectral functions in strongly anharmonic materials using CuI as example. 
\end{abstract}

\pacs{Valid PACS appear here}
\maketitle

\section{Introduction}
The electronic band structure is a fundamental concept in solid state physics and material science. Many key characteristics of crystalline materials, including transport, optical, magnetic and topological properties, can be qualitatively understood or even quantitatively computed using the band structure~\cite{martin_electronic_structure}. Over the past few decades, the development of density-functional theory~(DFT)~\cite{kohn_sham_dft,engel_dft_book} has enabled predictive calculations of electronic band structures from first principles. To qualitatively evaluate the properties of a simple, defect-free material with DFT, only calculations in the primitive cell (PC) crystal structure are needed for obtaining the band structure. However, to accurately capture material properties under more realistic conditions such as thermal vibrations~\cite{Knoop2020,Zacharias2020,Hellman2011}, defects~\cite{defect_example_1,defect_example_2}, interfaces~\cite{interface_example_1,interface_example_2} and doping~\cite{doping_example_1,doping_example_2}, large supercells (SC) that break the translational symmetry of the primitive cell are usually needed. The band structures calculated using these large SCs are significantly more difficult to interpret, since the volume of the first Brillouin zone (BZ) is inversely proportional to the real-space cell volume. As a result, the bands in the primitive cell Brillouin zone (PC-BZ) are folded into a much smaller supercell Brillouin zone (SC-BZ), causing many states to become indistinguishable. Additionally, this folding complicates a direct comparison with experimental results such as angle resolved photoemission spectroscopy (ARPES)~\cite{arpes_rmp_ZXShen}, which usually probes the electronic spectral function within the first PC-BZ~\cite{Xian.2022, arpes_rmp_ZXShen,suhuai_review}. To address these challenges,  band unfolding techniques can be employed, which map electronic states from the SC-BZ back to the PC-BZ. By this means, it is possible to shed light on the underlying physics of the symmetry-breaking perturbations,~e.g.,~vibrations, dopants, and defects, and analyze their influence on the band structure. In turn this facilitates the interpretation of the calculations and allows for a comparison with experimental observations~\cite{Allen2013,suhuai_review}.

The essential step in band unfolding is the projection of electron eigenvectors of the SC onto $\bf k$-points compatible with the PC translational symmetry. Hereby, the choice of projection strategy is dictated by the underlying
basis set representing the electron eigenvectors.  For example, many band unfolding implementations 
have been developed for plane-wave basis sets~\cite{zunger_bzu_pw,Bandup_bzu_pw,rubel_bzu_pw,mingxing_layer_bzu_pw,Fawei_Zheng_bzu_pw,vasp_bzu_pw}. This is relatively straightforward, since plane waves are orthogonal and unaffected by atomic displacements or defects.
Within a linear combination of atomic orbitals~(LCAO) ansatz, the actual basis functions are atom-centered,~i.e,~they are inherently 
tied to the atomic positions and species. Furthermore, LCAO basis functions are typically non-orthogonal, so the overlap matrix needs 
to be accounted for during unfolding. Existing approaches have so far circumvented these two hurdles,~e.g.,~by performing a Wannierization~\cite{ku_bzu_wannier,Wanniertools} or by representing the AOs in terms of plane waves, either implicitly~\cite{mayo_bzu_ao} or explicitly~\cite{Lixin_He_bzu_NAO}. While such plane-wave strategies are practically viable for the valence states typically addressed in pseudopotential-based calculations, this approaches become increasingly challenging for (semi-)localized core and semi-valence states, for which an immense amount of plane waves would be required. 
Consequently, existing band unfolding methods for LCAO basis sets are implemented in pseudopotential DFT codes~\cite{lee_bzu_LCAO,mayo_bzu_ao,Lixin_He_bzu_NAO}.
However, an efficient and accurate band unfolding methodology that is compatible with all-electron, full-potential DFT codes is still missing and hence urgently  needed.

In this work, we develop a new, efficient band unfolding algorithm for non-orthogonal LCAO basis sets and implement it in the all-electron, numeric atom-centered orbitals (NAOs) based {\it ab initio} materials simulation package \texttt{FHI-aims}~\cite{Blum2009,aims_roadmap}. To demonstrate the scalability of our implementation, we apply it to a band unfolding calculation of a large, 4,096-atom zinc blende GaN supercell comprising nearly 100,000 basis functions. Furthermore, we use the developed approach to compute non-perturbative temperature-dependent electronic spectral functions~\cite{Zacharias2020,nery_spectralfunction_feynmandiagram} for zinc blende CuI, a material known for its large anharmonicity at room temperature~\cite{florian_prl_2023,Knoop.2023}.

The paper is organized as follows. The derivation and implementation of the proposed band-unfolding approach is presented in Sec.~\ref{sec: methodology}. In particular, Sec.~\ref{sec: general theory} introduces the underlying, fundamental concepts, Sec.~\ref{sec: unfolding weight in LCAO} discusses the consequences for a non-orthogonal LCAO basis set, and Sec.~\ref{sec: analytical T_PC} comprises the analytical expressions needed for this representation. In Sec.~\ref{Implementation} and~\ref{sec: spectral_function}, the practical implementation details and the temperature-dependent electronic spectral function are introduced, respectively.  Sec.~\ref{sec: application} then covers an example application and  Sec.~\ref{sec: conclusion} summarizes our results.

\section{Methodology}
\label{sec: methodology}

\subsection{General Theory of Band Unfolding}
\label{sec: general theory}
In the following, we discuss band structure unfolding by establishing relationships between electronic states in the PC and those in a SC,~i.e.,~in a larger cell composed of multiple PCs. To distinguish between primitive and supercell quantities, we adopt a notation convention where lower-case letters represent PC-related quantities, and upper-case letters denote SC-related quantities. Additionally, PC and SC subscripts are used where necessary for clarity.
For instance, the three-dimensional lattice vectors of the primitive cell are denoted by $\mathbf{a}_{i} ~ (i=1,2,3)$, whereas those of the supercell are represented as $\mathbf{A}_{i}$. The associated lattice-vector matrix~$\matrixf{A}=[\mathbf{A}_{1},\mathbf{A}_{2},\mathbf{A}_{3}]$, in which each  $\mathbf{A}_{i}$ is a column vector, is related to the primitive lattice-vector matrix via the transformation:
\begin{equation}
\label{eq:transformation_matrix}
\matrixf{A}=\matrixf{a}\cdot\matrixf{M} \;.
\end{equation}
Here, $\matrixf{M}$ denotes a non-singular transformation matrix that consists of integer matrix elements~$M_{ij}$. Accordingly, the volume of the supercell is $m=\left|\text{det}(\mathbf{\underline{M}})\right|$ times larger than that of the PC.
Some important symbols are listed in Tab.~\ref{tab:symbols} for reference.

\begin{table*}
\centering
{
\begin{tabular}{cccc}
\hline \hline
Symbol & Description & Symbol & Description\\
\hline
$\matrixf{a} \ / \ \matrixf{A}$  & Lattice vector matrix of PC/SC & $ \matrixf{P}_{\bf k}$ & Projection operator to the eigenspace of $\bf k$ \\
$\matrixf{b} \ / \ \matrixf{B}$   & Reciprocal lattice vector matrix of PC/SC & $\ket{\varphi_{{\bf L}J}}$ & AO basis function of SC \\
${\matrixf{M}}$   & Transformation matrix between PC and SC & $\ket{\chi_{{\bf K}J}}$ & Bloch-type AO basis function of SC \\
$\matrixf{t} \ / \ \matrixf{T}$ & Translational operator of PC/SC & $\matrixf{\tilde{T}}_{\rm PC}$ & PC translational operator in the AO basis of SC \\
$\matrixf{h} \ / \ \matrixf{H}$ & Hamiltonian of PC/SC & $C_{{\bf k}n} \ / \ C_{{\bf K}N}$ & Expansion coefficient of $\ket{\psi_{{\bf k}n}} / \ket{\Psi_{{\bf K}N}}$ in  $\ket{\chi_{{\bf K}J}}$\\
${{\bf k}} \ / \ {{\bf K}}$ & Eigenvalues of PC/SC translational operator & $\matrixf{{T}}_{\rm PC}$ &  $\matrixf{\tilde{T}}_{\rm PC}$ after L{\"o}wdin transformation\\
${{\bf f_k}} \ / \ {{\bf F_K}}$ & Eigenvalues ${{\bf k}} / {{\bf K}}$ in the basis of $\matrixf{b} / \matrixf{B}$ & ${\bf F}_{{\bf k}n}$ & Eigenvectors of $\matrixf{T}_{\rm PC}$ \\
$\ket{{\bf k}j} \ / \ \ket{{\bf K}J}$ & Eigenvectors of PC/SC translational operator & ${\bf D_i}$ & Blocks of matrix $\matrixf{{T}}^{\bf K}_{\rm PC}$ in the SC $\bf K$ space\\
${\epsilon_{{\bf k}n}} \ / \ {E_{{\bf K}N}}$ & Eigenvalues of the PC/SC Hamilatonian & ${\bf V_i}$ & Eigenvectors of ${\bf D_i}$ \\
$\ket{\psi_{{\bf k}n}} \ / \ \ket{\Psi_{{\bf K}N}}$ & Eigenvectors of the PC/SC Hamilatonian \\
\hline
\end{tabular}
}
\caption{List of representative symbols used throughout this work.}
\label{tab:symbols}
\end{table*}

To solve the eigenvalue problem $\matrixf{h}\ket{\psi}=\epsilon\ket{\psi}$ viz. $\matrixf{H}\ket{\Psi}=E\ket{\Psi}$ in the PC and SC, periodic boundary conditions~(PBC) and the Bloch theorem are employed. This allows the definiton of the translational operators $\matrixf{t}_i$ and $\matrixf{T}_i$, which correspond to a translation of the whole system by one lattice vector~$\mathbf{a}_i$ and~$\mathbf{A}_i$, respectively. Since the commutators $[\matrixf{h},\matrixf{t}_i]=[\matrixf{H},\matrixf{T}_i]=0$ hold for all $i$, 
the solutions $\ket{\psi}/\ket{\Psi}$ of the respective Schrödinger equations can be expanded in terms of the eigenvectors of the translational operators, denoted as $\ket{{\bf k}j}$ / $\ket{{\bf K}J}$. Here, $j / J$ enumerate the degenerate eigenvectors corresponding to the eigenvalue $\bf k/K$ of the translational operator $\matrixf{t} / \matrixf{T}$. with this, we obtain:
\begin{eqnarray}
\ket{\psi_{{\bf k}n}} & = & \sum_{j} c_{{\bf k}n,j}  \ket{{\bf k}j} \\
\ket{\Psi_{{\bf K}N}} & = & \sum_{J} C_{{\bf K}N,J}  \ket{{\bf K}J} \label{eq:wf_sc}\;.
\end{eqnarray}
The components $c_{{\bf k}n,j}$ and $C_{{\bf K}N,J}$ are determined by solving the Schr{\"o}dinger equations $\matrixf{h}_{\bf k}\ket{\psi_{{\bf k}n}}=\epsilon_{{\bf k}n}\ket{\psi_{{\bf k}n}}$ viz. $\matrixf{H}_{\bf K}\ket{\psi_{{\bf K}N}}=E_{{\bf K}N}\ket{\psi_{{\bf K}N}}$, in which the Hamiltonian in the subspace $\bf k / K$ are given by the projection operator:
\begin{equation}
\label{eq: projector linalg}
    \matrixf{P}_{\bf k} = \sum_j \ket{{\bf k}j}\bra{{\bf k}j} \;,
\end{equation}
with $\matrixf{h}_{\bf k}=\matrixf{P}_{\bf k}\matrixf{h}\matrixf{P}_{\bf k}$ and $\matrixf{H}_{\bf K}=\matrixf{P}_{\bf K}\matrixf{H}\matrixf{P}_{\bf K}$.
The required eigenvectors of the translational operator are obtained by solving the eigenvalue problems $\matrixf{t}\ket{{\bf k}j} = \exp(i \vec{k} \cdot \vec{a}) \ket{{\bf k}j}$ and $\matrixf{T}\ket{{\bf K}J} = \exp(i \vec{K} \cdot \vec{A}) \ket{{\bf K}J}$. As mentioned above, $j / J$ is used because the eigenvectors of the translational operator can be degenerate and hence span an eigenspace~\cite{done_right}. 

Since the lattice vectors appear in the eigenvalues, the corresponding exponential eigenvalues exhibit different periodicities.  This is reflected in the periodicity of the reciprocal-lattice matrices that span the PC-BZ~$\matrixf{b}$ and the reduced SC-BZ~$\matrixf{B}$. They are defined via $\matrixf{b}\cdot\matrixf{a}=\matrixf{B}\cdot\matrixf{A}=2\pi\matrixf{I}$, where $\matrixf{I}$ is the identity matrix and the individual reciprocal lattice vectors $\mathbf{b}_{i}$ viz.~$\mathbf{B}_{i}$ are rows of $\matrixf{b}$ viz. $\matrixf{B}$. Accordingly, the SC-BZ and PC-BZ are related by:
\begin{equation}
    \label{eq: reciprocal lattice vector}
    \matrixf{B}=\matrixf{M}^{-1}\matrixf{b} \ .
\end{equation} 
This implies that the volume of the SC-BZ is $m$ times smaller than the PC-BZ. In other words, $m$ different eigenvalues $\vec{k}$ of the first Brillouin zone are mapped to one $\vec{K}$ in the reduced BZ, an effect commonly referred to as BZ folding. 

Eq.~(\ref{eq:transformation_matrix}) and Eq.~(\ref{eq: reciprocal lattice vector}) above describe basis transformations between the PC and the SC lattices. In practice, the corresponding eigenvalues ${\bf k}/{\bf K}$ are typically represented in fractional coordinates (${\bf f_k}$/${\bf F_K}$) within their respective reciprocal basis:
\begin{eqnarray}
    {\bf k} = {\bf f_k} \cdot \matrixf{b} {\quad {\rm viz.} \quad }
    {\bf K} = {\bf F_K} \cdot \matrixf{B} \ .
\end{eqnarray}
Accordingly, for one PC ${\bf k}$-point in the basis $\matrixf{b}$, the corresponding fractional coordinates ${\bf F_K}$ in the SC reciprocal basis $\matrixf{B}$ is given by:
\begin{eqnarray}
    \label{eq: fractional coords trans}
    {\bf F_K} = {\bf f_k} \cdot \matrixf{M} \ .
\end{eqnarray}
Eq.~(\ref{eq: fractional coords trans}) will be used to determine the SC $\bf K$-point coordinates from a given PC $\bf k$-point that we aim to unfold back to.
Note that the fractional coordinate ${\bf F}_{\bf K}$ calculated above can exceed 1.
However, all the ${\bf F}_{\bf K}$ and ${\bf F}_{\bf K'}$ that fulfill the relation ${\bf F}_{\bf K} = {\bf F}_{\bf K'} + p {\bf B}$ are equivalent due to the translational symmetry.
To remove this redundancy, we need to move all ${\bf F}_{\bf K}$ into the first BZ of SC via modulo operations.
Conversely, for one given $\bf K$-point in the reduced SC-BZ, the corresponding $m$ $\bf k$-points in the PC-BZ are linked by the relation:
\begin{equation}
    {\bf k}_p = {\bf K} + p{\bf B},
\end{equation}
where $p$ runs over all $m$ combinations of SC-BZ reciprocal-lattice vectors that can map $\bf K$ within the larger PC-BZ $\matrixf{b}$. 

\begin{figure}
    \centering
    \includegraphics[width=1\linewidth]{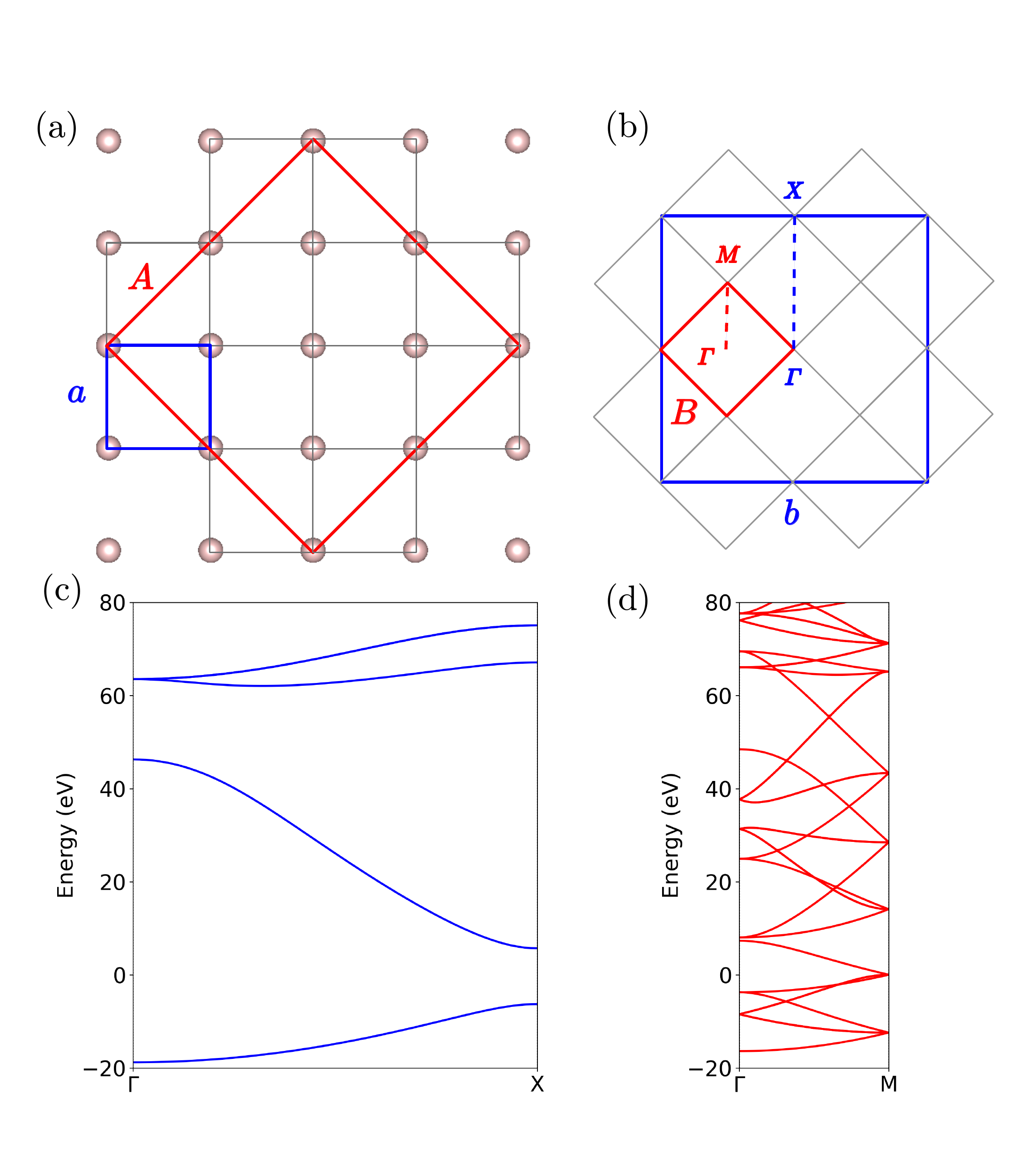}
    \caption{Atomic structure (a) and the first Brillouin zone (b) of a one-atom cubic hydrogen cell with lattice vector {$1.5$\AA} and a rotated, eight-atom SC. Blue square indicate the PC, red square indicate the SC and grey squares indicate replicas. (c) Band structure of the PC. (d) Band structure of the SC.}
    \label{fig:H_example}
\end{figure}

For example, Fig.~\ref{fig:H_example}(a) illustrates a one-atom cubic PC with a lattice vector of {$1.5$\AA} and a rotated, eight-atom SC obtained via the transformation matrix:
\begin{eqnarray}\label{eq:transformation_matrix_example}
    \matrixf{M} = 
    \begin{pmatrix}
        2 & 2 & 0 \\
        2 & -2 & 0 \\
        0 & 0 & 1 \\
    \end{pmatrix}.
\end{eqnarray}
As a result, the SC-BZ is eight times smaller than the PC-BZ, meaning that bands from eight PC $\bf k$-points are folded into one single $\bf K$-point in this small SC-BZ. The band folding effect significantly complicates the SC band structure, as evident in the comparison of Fig.~\ref{fig:H_example}(c) and (d), which makes a direct interpretation of the SC band structure more challenging. Note that, for a SC with non-diagonal $\matrixf{M}$, also the definition of $\bf k$-paths in SC/PC change after folding due to the change of symmetry. In this case, the original $\Gamma $(0,0,0) - X(0.5,0,0) path in the PC transforms into the $-$M($-0.5$,$-0.5$,0) - M(0.5,0.5,0) path in the SC, as shown in Fig.~\ref{fig:H_example}(b), which can be verified using Eq.~(\ref{eq: fractional coords trans}).

To recover the dispersion in the PC-BZ from the SC-BZ, it is hence necessary to project back the components of the SC wavefunction onto the eigenvectors~$\ket{{\bf k}j}$ of the translational operators associated with the PC. This is achieved using the projection operator $\matrixf{P}_{\bf k}$ corresponding to a given $\bf k$. The unfolding weights associated to each electronic state $\ket{\Psi_{{\bf K}N}}$ with energy $E_{{\bf K}N}$ are given by:
\begin{eqnarray}
\label{eq:unfolding_weight}
    W^{\bf k}_{{\bf K}N} & = & \bra{\Psi_{{\bf K}N}} \matrixf{P}_{\bf k} \ket{\Psi_{{\bf K}N}} \nonumber \\
                    & = & \sum_j |\braket{{\bf k}j|\Psi_{{\bf K}N}}|^2.
\end{eqnarray}
The projected basis $\ket{{\bf k}j}$ in the subspace ${\bf k}$ are usually chosen to be the PC eigenvector $\ket{\psi_{{\bf k}n}}$ hence the weights are expressed as:
\begin{eqnarray}
    W^{\bf k}_{{\bf K}N}
                    & = & \sum_n |\braket{\psi_{{\bf k}n}|\Psi_{{\bf K}N}}|^2.
\end{eqnarray}
Note that, $\ket{{\bf k}j}$ and $\ket{\psi_{{\bf k}n}}$ both form a complete basis of the subspace ${\bf k}$. Consequently, the projection $\matrixf{P}_{\bf k}$ is equivalent no matter which basis is used.
For a ``perfect'' SC,~i.e.,~one composed of identical PC replicas, these weights are strictly $0$ or $1$. This follows from the commutation relations 
$[\matrixf{H},\matrixf{t}_i]=[\matrixf{H},\matrixf{T}_i]=0$, which hold for all directions $i$ because the real-space Hamiltonians~$\matrixf{h}=\matrixf{H}$ are identical 
for the PC and the SC. However, in the case of a perturbed SC, where $\matrixf{H}\neq\matrixf{h}$, the PC translational operators $\matrixf{t}_i$ no longer commute with $\matrixf{H}$, and this condition no longer holds. 
Accordingly, the weights $W^{\bf k}_{{\bf K}N}$ become fractional. This enables the definition of a density-of-state-like spectral function as function of the energy~$E$:
\begin{eqnarray} \label{eq:spectral function}
    A({\bf k}, E) &=& \sum_{KN} W^{\bf k}_{{\bf K}N}  \delta (E - E_{{\bf K}N}) \;,
\end{eqnarray}
which will be discussed in more detail in Sec.~\ref{sec: spectral_function}.

Alternatively, band unfolding can also be understood from a group theory perspective. 
For a system that is cyclically periodic with a period of $l$ PCs, all possible translations of the PCs form a translation group ${\bf G}_{\bf PC}$ of order $l$, with $l$ characters of the irreducible representations labeled by $\bf k$. 
For the same system, translations by a SC that is $m$-times larger than the PC form another translation group ${\bf G}_{\rm SC}$ with order $l/m$ that is associated with $l/m$ characters of irreducible representations labeled by $\bf K$. This group is a subgroup of ${\bf G}_{\rm PC}$. 
Using ${\bf G}_{\rm SC}$ as the kernel of a homomorphism with ${\bf G}_{\bf PC}$, we can define the corresponding quotient group ${\bf G_Q} = {\bf G}_{\bf PC}/{\bf G}_{\bf SC}$, which has order $m$. Since translation groups are Abelian, the PC translation group can be expressed as a direct product decomposition~\cite{group_theory_unfolding_huang}:
\begin{eqnarray}
    {\bf G}_{\bf PC} = {\bf G}_{\bf SC} \otimes {\bf G_Q} \ ,
\end{eqnarray}
where the $m$-order translation group ${\bf G_Q}$ represents the $m$ possible translations $\hat{\bf t}_i$ of PCs within a single $m$-times larger SC. Equivalently, we can also generate a coset representation of ${\bf G}_{\bf PC}$ using ${\bf G}_{\bf SC}$ and the group elements in ${\bf G}_{\bf Q}$~\cite{ikeda2017mode}. As a result, a single character $\bf K$ of ${\bf G}_{\bf SC}$ corresponds to $m$ characters $\bf k$ of ${\bf G}_{\bf PC}$, which can be obtained as a direct product with ${\bf G}_{\bf Q}$.

Therefore, the projection operator onto $\bf k$ can also be constructed using the translational operations in ${\bf G}_{\bf Q}$ as~\cite{Allen2013,ikeda2017mode}:
\begin{eqnarray}
\label{eq: projector group theory}
    \matrixf{P}_{\bf k}
    &=& \frac{1}{m}\sum^{m}_{i} e^{-i{\bf k}\cdot {\bf l}_i} \ \hat{\bf t}_i \ ,
\end{eqnarray}
where $i$ indexes the $m$ PC translations $\hat{\bf t}_i$ within the SC, and ${\bf l}_i$ denotes the corresponding lattice translation. The factor $\exp (-i{\bf k}\cdot{{\bf l}_i})$ is actually the complex conjugate of the character of $\hat{\bf t}_i$.
This form of projection operator is commonly employed to construct symmetry-adapted basis functions corresponding to an irreducible representation in group theory~\cite{dresselhaus_group_theory}, and it is equivalent to the definition given in Eq.~(\ref{eq: projector linalg}), as shown in Appendix.~\ref{append: projector}. 
Analogously, the state-resolved unfolding can be performed by constructing projection operators for all irreducible representations of the little group at $\bf k$ using the corresponding point group operations~\cite{ikeda2017mode,dresselhaus_group_theory}.

\subsection{Band Unfolding in a Non-Orthogonal LCAO Basis Set}
\label{sec: unfolding weight in LCAO}
Starting from the general relations for band structure unfolding discussed in the previous section, we here develop practical formulas for application in electronic-structure theory. These expressions are more involved, though, since finite basis sets are used to numerically represent the wave functions~$\ket{\psi}$ and~$\ket{\Psi}$. In this work, we develop an efficient band unfolding approach tailored for non-orthogonal LCAO basis sets. 

In the LCAO ansatz, the electron
wave functions are expanded by a set of AOs. In our {\tt FHI-aims}
implementation, these are NAOs, in which {the radial functions are numerically tabulated.}
Such (numeric) atomic orbitals in real space are typically denoted as $\aoeq{{\bf L}}{J}
\equiv \varphi_J ({\bf r} - {\bf L})$, highlighting that the orbital is centered at the corresponding atomic position belonging to the cell index {${\bf L} = (L_1A_1, L_2A_2, L_3A_3)$ in three dimensions} 
and the orbital index inside each cell is $J$.
In general, AO basis are non-orthogonal, so that the overlap matrix $\matrixf{S}_{{\bf L}I,{\bf L'}J}=\braket{\varphi_{{\bf L}I}|\varphi_{{\bf L'}J}}$ 
is not diagonal~($\matrixf{S} \neq \matrixf{I}$).

{Due to the periodicity of crystal structures,  the wavefunction needs to satisfy the Bloch theorem: $\matrixf{\tilde{T}} \ket{\Psi_{{\bf K}N}} = e^{i{\bf K\cdot A}} \ket{\Psi_{{\bf K}N}}$, where $\matrixf{\tilde{T}}$ is the translational operator in AO basis, which moves each AO into the previous unit cell (${\bf L-1}$):
\begin{equation} \label{eq: translation}
    \matrixf{\tilde{T}} \aoeq{{\bf L}}{J}
= \aoeq{({\bf L-1})}{J} .
\end{equation}
This implies that the wavefunction must lie in the eigenspace of $\matrixf{\tilde{T}}$ corresponding to each eigenvalue $\bf K$. From the definition (\ref{eq: translation}) of $\matrixf{\tilde{T}}$ in an AO basis, a simple basis $\blochaoeq{{\bf K}}{J}$ for the eigenspace $\bf K$,~i.e.,~$\ket{{\bf K}J}$ in Eq.~(\ref{eq:wf_sc}), can be constructed as:
\begin{equation} \label{eq:ao to bao}
    \ket{\chi_{{\bf K}J}} = \frac{1}{\sqrt{L_{tot}}} \sum_{\bf L} e^{i{\bf K\cdot L}} \aoeq{{\bf L}}{J} .
\end{equation}
Note that the above expression directly follows from the definition of $\matrixf{\tilde{T}}$ given in Eq.~(\ref{eq: translation})
and holds for orthogonal and non-orthogonal AOs. 
}

Accordingly, each energy eigenstate $\ket{\Psi_{{\bf K}N}}$ with band index $N$ can be expressed as a
linear combination of such Bloch-type AOs:
\begin{equation} \label{eq:bao to wf}
    \ket{\Psi_{{\bf K}N}} = \sum_J C_{{{\bf K}N},J} \blochaoeq{{\bf K}}{J} \ .
\end{equation}
Due to the non-orthogonality of these AOs, the electron wavefunctions are determined by a generalized eigenvalue problem in reciprocal space:
\begin{equation} \label{eq: gev}
    \matrixf{H}_{\bf K}\mathbf{C}_{{\bf K}N} = E_{{\bf K}N} \matrixf{S}_{\bf K} \mathbf{C}_{{\bf K}N} \ ,
\end{equation}
where $\mathbf{C}_{{\bf K}N}$ is the expansion coefficient vector of $\ket{\Psi_{{\bf K}N}}$ in a Bloch-type
AO basis \{$\ket{\chi_{{\bf K}J}}$\}. We will refer to $\mathbf{C}_{{\bf K}N}$ also as ``electron wavefunction" in the following. Furthermore, $\matrixf{S}_{\bf K}$ is the overlap matrix in $\bf K$ space and features the matrix elements:
\begin{equation}
    \matrixf{S}_{{\bf K}I,{\bf K}J} = \langle \chi_{\tmop{{\bf K}I}}\blochaoeq{{\bf K}}{J} = \sum_{\bf L} e^{\tmop{i{\bf K\cdot L}}} \matrixf{S}_{0I,{\bf L}J} \ .
\end{equation}
The orthonormalization condition for $\mathbf{C}_{{\bf K}N}$ is given by:
\begin{eqnarray} \label{nonortho normalization}
    \langle \Psi_{{\bf K}M}|\Psi_{{\bf K}N} \rangle &=& \sum_{IJ}C^{\ast}_{{\bf K}M, I} \langle \chi_{{\bf K}I}\ket{\chi_{{\bf K}J}} C_{{\bf K}N, J} \nonumber \\
    &=& \mathbf{C}^{\dagger}_{{\bf K}M}\matrixf{S}_{\bf K}\mathbf{C}_{{\bf K}N} = \delta_{{MN}} \ .
\end{eqnarray}
{A detailed discussion of the representation of matrix elements and eigenvalue equations in a non-orthogonal basis is provided in Appendix.~\ref{appendix_matrix_non_ortho}. 

In a LCAO code
implementation, the unfolding weights can thus be expressed in terms of $\mathbf{C}_{{\bf K}N}$ as:
\begin{eqnarray} \label{eq:unfold nonorthogonal}
    W^{\bf k}_{{\bf K}N}
                    & = & \sum_n |\braket{\psi_{{\bf k}n}|\Psi_{{\bf K}N}}|^2 \nonumber
    \\
    &=& \sum_n \sum_{I J} 
    |C^{\ast}_{{\bf k}n, I} \langle \chi_{{\bf K}I}\ket{\chi_{{\bf K}J}} C_{{\bf K}N, J}|^2 \nonumber
    \\ 
    &=& \sum_n |\mathbf{C}^{\dagger}_{{\bf k}n}\matrixf{S}_{\bf K}\mathbf{C}_{{\bf K}N}|^2 \ .
\end{eqnarray}
The SC wavefunction $\mathbf{C}_{{\bf K}N}$ is obtained from the band structure calculation in the SC. To evaluate Eq.~(\ref{eq:unfold nonorthogonal}) via matrix multiplication, we also need to express the PC wavefunction $\mathbf{C}_{{\bf k}n}$ in the SC basis. In principle, this can be achieved by directly calculating the eigenvectors of the PC
hamiltonian $\matrixf{h}$ and then transforming them into the SC basis. 
In a plane-wave basis code, this transformation is straightforward since the basis functions are unrelated to the atom positions or species. However, in an LCAO basis code, this is much more complicated since the basis functions are now directly associated with the atomic positions {and species}. {When the atoms in the SC are perturbed, the basis functions themselves change, adding another layer of complexity.}
To address this problem, we adopt a different approach which does not require the explicit calculation of the PC wavefunction. {Instead, this method allows us to work directly} in the SC $\bf K$-space, 
avoiding the transformation of wavefunctions between the PC and SC basis.

From the definition of the unfolding weights in Eq.~(\ref{eq:unfolding_weight}), the key to the band unfolding problem is determining the projection operator $\matrixf{P}_{\bf k}$. 
Any complete basis set $\ket{{\bf k}j}$ within the eigenspace ${\bf k}$ can be used for this purpose. This implies that if we can obtain the matrix representation of $\tmnao_{PC}$ in the SC basis and identify any set of its eigenstates $\{\ket{\xi_{kn}}\}$, then we can perform the projection into the subspace ${\bf k}$ using $\{\ket{\xi_{kn}}\}$ instead:
\begin{eqnarray}
\label{eq:unfolding_weight_in_nao}
    W^{\bf k}_{{\bf K}N}
                    & = & \sum_n |\braket{\psi_{{\bf k}n}|\Psi_{{\bf K}N}}|^2 \nonumber = \sum_n |\langle\xi_{{\bf k}n}\ket{\Psi_{{\bf K}N}}|^2 \\
     &=& \sum_n |\tilde{\mathbf{F}}_{{\bf k}n}^{\dagger}\matrixf{S}_{\bf K}\mathbf{C}_{{\bf K}N}|^2 \ .
\end{eqnarray}
Here, $\tilde{\mathbf{F}}_{{\bf k}n}$ are expansion coefficients of the eigenvectors $\ket{\xi_{{\bf k}n}}$ in the SC basis. 
The eigenvectors $\{\ket{\xi_{{\bf k}n}}\}$ are obtained by solving the eigenvalue problem $\tmnao_{PC}^{\bf K}\ket{\xi_{{\bf k}n}}=e^{i{\bf k\cdot a}}\ket{\xi_{{\bf k}n}}$ at each SC $\bf K$ point. The set of eigenvalues contains exactly those PC $\bf k$ points to which this SC $\bf K$ point is mapped. Note that the eigenvectors~$\ket{\xi_{{\bf k}n}}$~ of $\tmnao_{PC}$ may not coincide with the eigenvectors~$\ket{\psi_{{\bf k}n}}$ of $\matrixf{h}$, even though they commute, since the $\tmnao_{PC}$ can have higher degeneracy~\cite{CohenTannoudji}. However, for calculating the unfolding weights, any set of eigenvectors is equivalent as mentioned above.
By this means, the unfolding problem is reduced to finding the matrix representation of $\tmnao_{PC}$ and diagonalizing it in the SC basis.

Now, let us consider the general matrix form of $\tmnao$ in a non-orthogonal basis. According to Eq.~(\ref{eq: translation}), $\tmnao$ shifts AOs to the previous lattice sites. Here, we define $\braket{\varphi_{{\bf L'}I}|\tmnao|\varphi_{{\bf L}J}}$ as the matrix elements of  $\tmnao$ as discussed in Appendix.~\ref{appendix_matrix_non_ortho}. Consequently, the matrix elements that enter the eigenvalue equation are:
\begin{equation}
    \langle \varphi_{{\bf L'}I}|\tmnao\ket{\varphi_{{\bf L}J}}
    = \braket{\varphi_{{\bf L'}I} | \varphi_{({\bf L - 1})J}}
    = \mathbf{S}_{{\bf L'}I,({\bf L - 1})J} \ .
\end{equation}
This shows that the real-space matrix element $\langle \varphi_{{\bf L'}I}|\tmnao\ket{\varphi_{{\bf L}J}}$ in a non-orthogonal basis corresponds to overlap matrix element of basis functions in neighboring cells. As shown in Fig.~\ref{fig:T_matrix_nao}, the translational operator in a non-orthogonal basis actually transforms the overlap matrix. From this relation, we observe that in a non-orthogonal AO
basis $\tmnao = \matrixf{S}\matrixf{T}$, where $\matrixf{T}$ is the permutation matrix, i.e., the corresponding translational matrix in an orthogonal basis. Furthermore, $\matrixf{T}$ commutes with $\matrixf{S}$,~i.e.,~$[\matrixf{T}, \matrixf{S}] = 0$ under periodic boundary conditions. 
Here, we prove this using one AO in each unit cell, but the proof extends naturally to multiple AOs per unit cell. By definition, $\matrixf{ST}$ shifts each column of $\matrixf{S}$ to the previous column and  $\matrixf{TS}$ shifts each
line of $\matrixf{S}$ to the next line, following the standard matrix transformation rule. Thus, the matrix elements satisfy:  $$(\matrixf{ST})_{\tmop{ij}} =
\matrixf{S}_{i (j - 1)} = \matrixf{S}_{(i + 1) j} = (\matrixf{TS})_{\tmop{ij}} \ ,$$
where we have used the
translational symmetry of the overlap matrix $\matrixf{S}_{(i + n) (j + n)} =
\matrixf{S}_{{ij}}$.

\begin{figure}[t]
	\center{\includegraphics[width=1\linewidth]{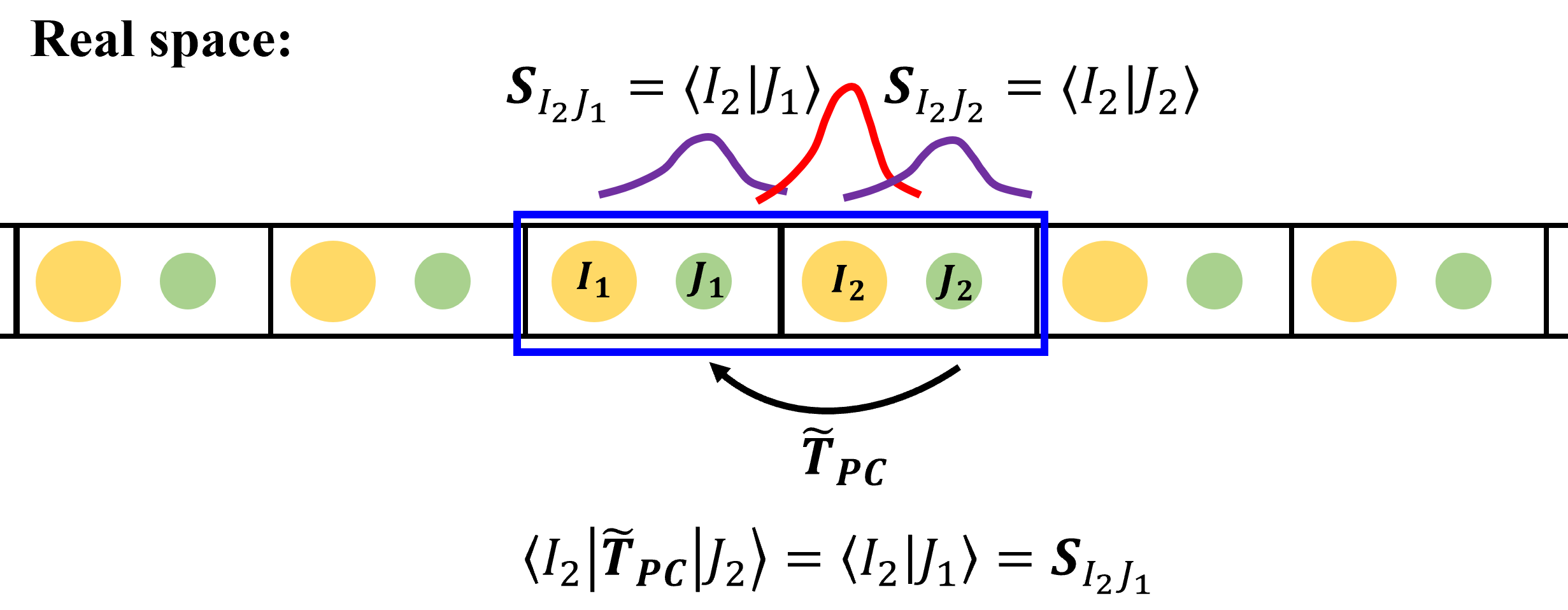}}
	\caption[]{Illustration of the action of the translational operator in a non-orthogonal LCAO basis. Red and purple curves represent the AOs $I$ and $J$.}
	\label{fig:T_matrix_nao}
\end{figure}

After defining the translational matrix, we can further simplify the problem by performing a L\"{o}wdin transformation, which eliminates the overlap matrix in the unfolding weights in Eq.~(\ref{eq:unfolding_weight_in_nao}). By defining:
\begin{eqnarray}
    \mathbf{C}'_{{\bf K}N} &=& \mathbf{S}^{1 / 2} \mathbf{C}_{{\bf K}N} \\
    \matrixf{H}_{\bf K}' &=& \matrixf{S}_{\bf K}^{- 1 /2} \matrixf{H}_{\bf K}\matrixf{S}_{\bf K}^{- 1 / 2} \ ,
\end{eqnarray}
we turn Eq.~(\ref{eq: gev}) into a standard eigenvalue equation: $\mathbf{H}_{\bf K}' \mathbf{C}'_{{\bf K}N} = E_{{\bf K}N}
\mathbf{C}'_{{\bf K}N}$. The unfolding weight then becomes:
\begin{eqnarray}
    W^{\bf k}_{{\bf K}N} &=& \sum_n |\tilde{\mathbf{F}}_{{\bf k}n}^{\dagger}\mathbf{S}^{1/2}\mathbf{S}^{1/2}\mathbf{C}_{{\bf K}N}|^2 \nonumber \\
    &=& \sum_n |\tilde{\mathbf{F}}_{{\bf k}n}^{\dagger}\mathbf{S}^{1/2}\mathbf{C}'_{{\bf K}N}|^2 \ .
\end{eqnarray}
Similarly, we apply the L{\"o}wdin transform to the $\tmnao_{\tmop{PC}}^{\bf K}$
to obtain $\tilde{\mathbf{F}}'_{{\bf k}n} =\mathbf{S}^{1 / 2} \tilde{\mathbf{F}}_{{\bf k}n}$, which yields:
\begin{eqnarray}
    \tmnao_{\tmop{PC}}' &=& \matrixf{S}^{- 1 /
2} \tmnao_{\tmop{PC}} \matrixf{S}^{- 1 / 2} \nonumber \\ &=& \matrixf{S}^{- 1 /
2}\tmpc\matrixf{S}\matrixf{S}^{- 1 /
2} = \tmpc \ .
\end{eqnarray}
Here, we have omitted the index $\bf K$.
We also use the theorem that if $\matrixf{T}$ commutes with $\matrixf{S}$, then it also commutes with any function of $\matrixf{S}$. 

As demonstrated above, the L{\"o}wdin transformation makes $\tmnao_{\tmop{PC}}'$ identical to the translational matrix in an orthogonal basis. Consequently, the eigenvectors $\tilde{\mathbf{F}}'_{{\bf k}n}$ are simply the eigenvectors of $\tmpc$ in an orthogonal basis, i.e., $\tilde{\mathbf{F}}'_{{\bf k}n}=\mathbf{F}_{{\bf k}n}$, where $\mathbf{F}_{{\bf k}n}$ are eigenvectors of $\tmpc$. The unfolding weights in a non-orthogonal LCAO basis then simplify to:
\begin{equation} \label{eq:unfoldweight nao}
    W^{\bf k}_{{\bf K}N} = \sum_n |\mathbf{F}_{{\bf k}n}^{\dagger}\mathbf{C}'_{{\bf K}N}|^2 \ .
\end{equation}
This is helpful, since $\mathbf{F}_{{\bf k}n}$ can be expressed analytically, as shown in the next section.
This makes our method much faster and simpler.

\subsection{Analytical Expression of the Projection Operator}
\label{sec: analytical T_PC}
To compute the weights $W^{\bf k}_{{\bf K}N}$, we first need to determine the eigenvectors $\mathbf{F}_{{\bf k}n}$. To achieve this, we begin by analyzing the structure of the translational matrix through an example. 
We hence consider a system with two AOs per PC that is described via a Born-von-Karnman cell as cyclically periodic in real space with a period of $l$ PCs.  Accordingly, the system features $l$ discrete translations, corresponding to $l$ $\bf k$-points in reciprocal space. 
In this case, the $\tmpc$ in an orthogonal AO basis takes the form:
\begin{eqnarray}
    \tmpc = 
    \begin{pmatrix}
0 & 0 & 1 & 0 & 0 & 0 & \cdots & 0\\
0 & 0 & 0 & 1 & 0 & 0 & \cdots & 0 \\
0 & 0 & 0 & 0 & 1 & 0 & \cdots & 0 \\
0 & 0 & 0 & 0 & 0 & 1 & \cdots & 0 \\
\vdots & \vdots & \vdots & \vdots & \vdots & \vdots &\ddots & \vdots \\
0 & 0 & 0 & 0 & 0 & 0 & \cdots & 1\\
1 & 0 & 0 & 0 & 0 & 0 & \cdots & 0\\
0 & 1 & 0 & 0 & 0 & 0 & \cdots & 0\\
    \end{pmatrix}\;.
\end{eqnarray}
The dimension of matrix $\tmpc$ is $(2 l \times 2 l)$ and every two AOs belonging to the same unit
cell are moved to the corresponding positions in the neighboring cell. 
For example, the first row shows that the first AO in the first PC~(index one) is mapped to the position of the first AO in the second PC, i.e., the one with index three. 
This matrix is a real permutation matrix, in which the elements can only be 0 or 1 as mentioned above.

Now consider a SC consisting of $m$ PCs with $p$ AOs in each PC. Still, this ``perfect'' supercell retains the real-space periodicity of the previous example featuring $l$ PCs. Consequently, the number of translational operations consistent with the SC reduces to $l/m$, corresponding to $l/m$ distinct $\bf K$-points. This reduction is the origin of the band folding effect observed in supercell calculations.
In the SC $\bf K$-space, we thus obtain the following representation of $\tmpc$:
\begin{equation}
    \braket{\chi_{{\bf K}I}|\tmpc^{\bf K}|\chi_{{\bf K}J}} = \sum^{l/m}_{\bf L} e^{\tmop{i{\bf K\cdot L}}} \braket{\varphi_{0I}|\tmpc|\varphi_{{\bf L}J}} \ ,
\end{equation}
where $\bf L$ is the SC index and there are $l/m$ SCs in total. 
Unlike the real-space matrix discussed above, the elements of the ($pm \times pm$)-dimensional 
matrix $\tmpc^{\bf K}$ are not just 1 or 0, as they now carry complex phase factors introduced by the Fourier transformation.
Moreover, $\tmpc^{\bf K}$ can be recast into a block-diagonal form:
\begin{eqnarray}
    \tmpc^{\bf K} = 
    \begin{pmatrix}
     \mathbf{D_1} & 0 & 0 & \cdots & 0\\
     0 & \mathbf{D_2} & 0 & \cdots & 0\\
     0 & 0 & \mathbf{D_3} & \cdots & 0\\
     \vdots & \vdots & \vdots & \ddots & \vdots \\
     0 & 0 & 0 & \cdots & \mathbf{D_p}\\
    \end{pmatrix}\;,
    \label{matrix:TK_PC}
\end{eqnarray}
featuring $p$ blocks $\mathbf{D_i}$. These ($m \times m$)-dimension blocks take the form:
\begin{eqnarray}
    \mathbf{D_i} = 
    \begin{pmatrix}
     0 & d_1 & 0 & \cdots & 0 & 0\\
     0 & 0 & d_2 & \cdots & 0 & 0\\
     0 & 0 & 0 & \cdots & 0 & 0\\
     \vdots & \vdots & \vdots & \ddots & \vdots &\vdots\\
     0 & 0 & 0 & \cdots & 0 & d_{m-1}\\
     d_m & 0 & 0 & \cdots & 0 & 0\\
    \end{pmatrix} \;.
    \label{matrix:D}
\end{eqnarray}
In this recast representation, there are $p$ blocks and, each block $\mathbf{D_i}$ describes the translations of one specific PC-AO in the SC. In turn, it runs over all $m$ PCs and each row/column has exactly one non-zero element $d_j$. The elements $d$ can take different values, if the translation by a PC lattice vector retains the atom within the same SC,
$d_j = 1$; if the translation makes the atom cross the SC boundary to the next SC, it carries a phase factor $d_j = e^{i{\bf K\cdot L}}$. The $\mathbf{D_i}$ block therefore describes a PBC-closed path with $m$-steps for translating the PC-AO $i$ within the SC under $\tmpc^{\bf K}$. 
For example, in Fig.~\ref{fig:T_matrix}, we consider a SC comprising two PCs with each PC containing two basis functions. As shown in Fig.~\ref{fig:T_matrix}(b), applying $\matrixf{T}_{\rm PC}$ shifts basis $I_1$ 
to the next SC, while $I_2$ remains within the same SC. Consequently, their corresponding values for $d_j$ are $e^{i{\bf K}}$ and $1$, respectively. The ($2\times2$) block $D_I$ represents the full path of the PC basis $I$ in the SC.

Eventually, let us note that from a group theory point of view, considering all basis functions in one SC as a set, all $m$ basis functions $\varphi_i$ in the SC that are linked by the $m$ PC translations $\tmpc$ form an orbit. Each orbit is an invariant subset, and the block $\mathbf{D_i}$ actually describes such an orbit. The eigenvalues of the matrix $\mathbf{D}$ represent the characters of the irreducible representations of the corresponding translation group ${\bf G_Q}$ mentioned above. Meanwhile, the eigenvectors serve as the basis that spans these irreducible representations.

\begin{center}
	\begin{figure}[ht!]
		
		\center{\includegraphics[width=1\linewidth]{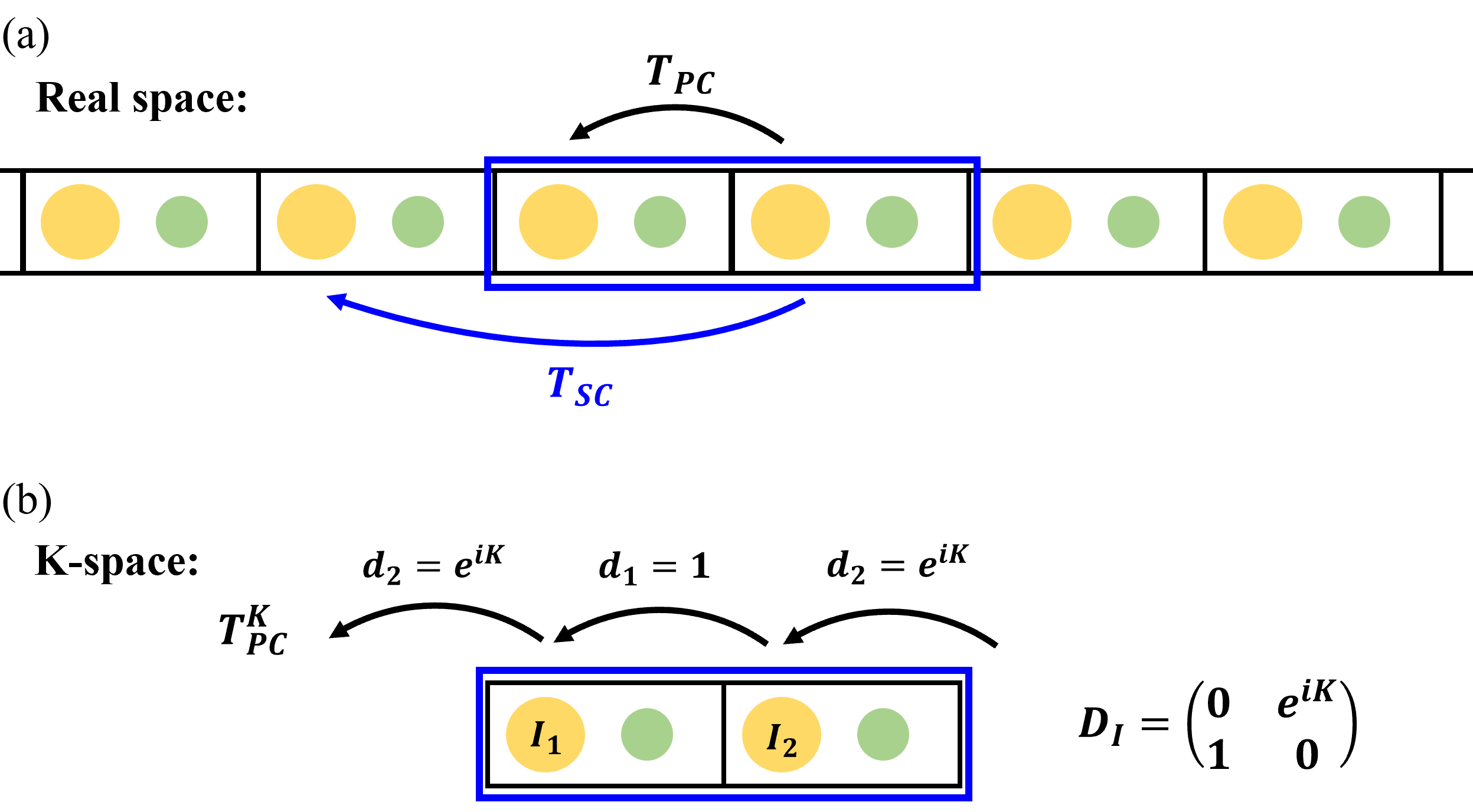}}
		
		\caption[]{(a) Illustration of how the translational operators acts on each AO in real space. Here we use the black lattice to represent our PC and the two AOs in the PC are represented by yellow and green circles. The SC consists of two PCs is represented 
    by the blue lattice. (b) Illustration of the matrix elements of the translational operator in K-space. We labeled the corresponding bloch-type AO belonging to two different PCs in our SC by $I_1$ and $I_2$. $D_I$ is the complex permutation matrix that
    represent the full path of orbital $I$ under the PC translational operator, as shown in Eq.(\ref{matrix:D}). }
		\label{fig:T_matrix}
	\end{figure}
\end{center}

In a last step, we need to find the eigenvectors of such matrices~$\mathbf{D_i}$.
For matrices with such a structure, the characteristic polynomial
can be calculated via the Leibniz formula:
\begin{equation}
    \lambda_l^m - e^{i \phi} \quad\text{with}\quad e^{i \phi} = \prod_{j=1}^m d_j \ .
\end{equation}
Thus, the $m$ eigenvalues of~$\mathbf{D_i}$  are given by:
\begin{equation}
    \lambda_l = \exp\left(i \frac{\phi + 2 \pi l}{m} \right) \quad l \in [0,\cdots, m) \ .
\end{equation}
Due to the structure of~$\mathbf{D_i}$ in Eq.~(\ref{matrix:D}), an analytical expression for the associated eigenvectors ${\bf V}_{l}=(V_{l 1},\cdots,V_{l m})$ can be expressed recursively in terms of~$V_{l 1}$:
\begin{eqnarray}
    V_{l 2} &=& \lambda_l V_{l 1} d_1^{\ast} = V_{l 1} \lambda_l d_1^{\ast} \\
    V_{l 3} &=& \lambda_l V_{l 2} d_2^{\ast} = V_{l 1} \lambda_l^2 d_1^{\ast}d_2^{\ast} \\
    &...& \nonumber \\
    V_{\tmop{lm}} &=& \lambda_l V_{l (m - 1)} d_{m - 1}^{\ast} = V_{l 1} \lambda_l^{m -
   1} d_1^{\ast} \ldots d_{m - 1}^{\ast}
\end{eqnarray}
Enforcing normalization allows for the determination of~$V_{l 1}$, enabling the
construction of  the eigenvectors $\mathbf{F}_{{\bf k}n}$ of $\tmpc^{\bf K}$. 
Let us note that
the eigenvalues $\lambda_l$ are identical for all the $p$ matrices~$\mathbf{D_i}$ 
associated with a specific~$\tmpc^{\bf K}$, 
as they undergo the same translation operation.
In other words, for each SC $\bf K$-point, there will be $m$ unfolded $\bf k$-points and $p$ eigenvectors at each $\bf k$. The $p$ eigenvectors $\mathbf{F}_{{\bf k}n}$ associated with the unfolded $\bf k_i$ can be constructed as:
\begin{equation}
\label{matrix:F_k}
     \matrixf{F}_{\bf k_i} = {\bf V_i} \otimes {\matrixf{I}_{p\times p}},
\end{equation}
or, more specifically, as:
\begin{eqnarray}
    \mathbf{F}_{{\bf k_i}1} = 
    \begin{pmatrix}
        {\bf V_i} \\
        0 \\
        \vdots \\
        0
    \end{pmatrix} , 
        \mathbf{F}_{{\bf k_i}2} = 
    \begin{pmatrix}
        0 \\
        {\bf V_i} \\
        \vdots \\
        0
    \end{pmatrix}, \cdots,
    \mathbf{F}_{{\bf k_i}p} = 
    \begin{pmatrix}
        0 \\
        0 \\
        \vdots \\
        {\bf V_i}
    \end{pmatrix}. \nonumber
\end{eqnarray}
With this analytical expression for $\mathbf{F}_{{\bf k}n}$, the unfolding weights can be efficiently calculated using Eq.~(\ref{eq:unfoldweight nao}). 

\begin{center}
    \begin{figure}
        \centering
        \includegraphics[width=1\linewidth]{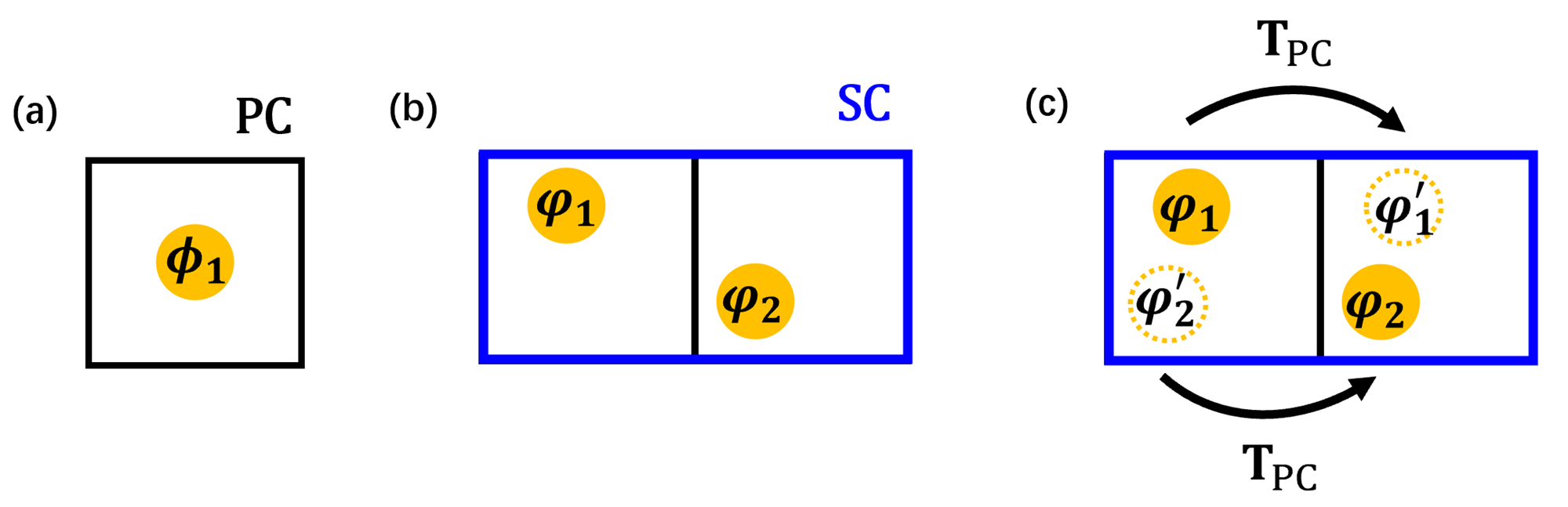}
        \caption{(a) The perfect PC with 1 atomic orbital (b) A distorted SC with displaced AOs.
        (c) The ``extended perfect" SC constructed by adding placeholder orbitals $\varphi'$ in the original distorted SC.}
        \label{fig:disturbed_sc}
    \end{figure}
\end{center}

Finally we note that the projection operator for SCs featuring displaced atoms can be constructed following exactly the same procedure described above for perfect PC replicas. This is rooted in the fact
that unfolding aims at projecting the SC states, even when these are associated to displaced atoms, onto a subspace spanned by projectors that fulfill the target PC symmetry.
In other words, we are not interested in state-specific weights~$W^{{\bf k}n}_{{\bf K}N} =|\braket{\psi_{{\bf k}n}|\Psi_{{\bf K}N}}|^2$ that depend on the LCAO representation of the individual PC eigenstates. Rather, we are computing the reciprocal-space mapping $\mathbf{k}\leftrightarrow \mathbf{K}$, i.e.,   $W^{{\bf k}}_{{\bf K}N} = \sum_n |\braket{\psi_{{\bf k}n}|\Psi_{{\bf K}N}}|^2$, which is fully defined by the lattice vectors of PC and SC.
In practice,
this is achieved by constructing translational operators that fulfill the PC symmetry also for those SC basis functions that do not. Let us exemplify this using a spinless, 1D system with 
lattice period $a$ and a single AO per unit cell, as illustrated in Fig.~\ref{fig:disturbed_sc}(a). The respective SC is twice as large and contain two AOs that explicitly break the PC 
symmetry, as shown in Fig.~\ref{fig:disturbed_sc}(b). In this model system, the SC wavefunctions can hence be expressed as:
\begin{eqnarray}
    \Psi_{{ K}N} &=& C_1 \chi_{{ K}1} + C_2 \chi_{{ K}2} \ , \\
    {\bf C}_{{ K}N} &=& (C_1, C_2)^T \ ,
\end{eqnarray}
where $\chi_{{ K}1}$ and $\chi_{{ K}2}$ are Bloch-type AOs constructed using $\varphi_1$ and $\varphi_2$. Due to the atomic distortions, the SC basis functions $\varphi_1$ and $\varphi_2$ are not connected by PC translations. Nonetheless, the projectors needed for unfolding are constructed by relating $\varphi_1$ and $\varphi_2$ by the translational operator of the perfect PC:
\begin{equation}
\tmpc^{K} = {\bf D} =
    \begin{pmatrix}
        0 & 1 \\
        e^{i{ K}{A}} & 0 
    \end{pmatrix} \;,
    \label{eq:TPC_oneAO}
\end{equation}
as if the SC basis functions $\varphi_1$ and $\varphi_2$ would fulfill such a higher-symmetry translation. 
By diagonalizing this matrix, we obtain the eigenvectors ${\bf F}_{{ k}_1}$ and ${\bf F}_{{ k}_2}$ and hence
the unfolding weights:
\begin{eqnarray}
    W^{{ k}_1}_{{ K}N} &=& |{\bf F}^\dagger_{{ k}_1} {\bf C}_{{ K}N}|^2 
    = \frac{1}{2}|C_1 + e^{-i{ K a}}C_2 |^2 \ , \nonumber \\
        W^{{ k}_2}_{{ K}N} &=& |{\bf F}^\dagger_{{ k}_2} {\bf  C}_{{ K}N}|^2
    = \frac{1}{2}|C_1 + e^{-i{ (K+B)a}}C_2 |^2 \ .
\label{eq:weight_oneAO}
\end{eqnarray}
This procedure is tantamount to constructing an ``extended'' AO basis set for the SC via placeholder orbitals~$\{\varphi'_1, \varphi'_2\}$ that formally re-establish the ideal ${\bf T}_{\rm PC}$-translational symmetry in the AO basis set, as shown in Fig.~\ref{fig:disturbed_sc}(c). In this ``extended'' basis set, the number of AOs increases to four and the respective SC wave function in this ``extended'' basis becomes:
\begin{eqnarray}
    \Psi_{{ K}N} &=& C_1 \chi_{{ K}1} + 0 \chi'_{{ K}1} + 0 \chi'_{{ K}2} + C_2 \chi_{{ K}2} \\
    {\bf C}'_{{ K}N} &=& (C_1, 0, 0, C_2)^T \ .
\end{eqnarray}
Since all orbitals in the distorted SC now posses perfect replicas connected by PC lattice translations, the matrix representation of $\tmpc'^{K}$ is
\begin{equation}
\label{matrix pc distort}
        \tmpc'^{K} = 
    \begin{pmatrix}
     {\bf D} & 0  \\
     0 & {\bf D} 
    \end{pmatrix} \;, 
\end{equation}
which is nothing else than a block-diagonal matrix filled with the $\tmpc^{K}$ matrix defined in Eq.~(\ref{eq:TPC_oneAO}) above. 
Using the fact that the placeholder Bloch-orbitals~$\chi'_{{ K}1}$ and $\chi'_{{ K}2}$ do not contribute to the SC wave function, it is
straightforward to verify that this yields the exact same unfolding weights $W^{{ k}}_{{ K}N} = |{\bf F}^\dagger_{{ k}} {\bf C}_{{ K}N}|^2$ 
presented in Eq.~(\ref{eq:weight_oneAO}) above. 
This highlights that the projectors do not depend on the actual distortion of the SC and all sum rules such as $\sum_{{\bf k}n} W^{{\bf k}n}_{{\bf K}N} = 1$ and $\sum_{N}W^{{\bf k}n}_{{\bf K}N} = 1$  are numerically fulfilled in our unfolding implementation even for displaced atoms.
The presented procedure can also be applied
in the case of vacancies, interstitials, and impurities by introducing placeholder orbitals as discussed above.

\subsection{Implementation Details}
\label{Implementation}
In a practical implementation of the formalism derived above, several additional details have to be considered. For instance, we have so far only considered a 1D translation. In a periodic 3D system, however, translations occur along all three Cartesian directions. Since these translations are Abelian, they can be diagonalized sequentially.
To this end, we first construct the translational operator for the first PC lattice vector $\matrixf{T}^{(1)}_{PC}$ and obtain its eigenvectors, denoted as $\matrixf{F}^{(1)}_{\bf k_1}$. Next, we can expand the corresponding $\matrixf{T}^{(2)}_{PC}$ for the second PC lattice vector in the eigenspace of $\matrixf{F}^{(1)}_{\bf k_1}$ via
\begin{equation}
\matrixf{T}^{\rm red (2)}_{PC} = \matrixf{F}^{\dagger(1)}_{\bf k_1}\matrixf{T}^{ (2)}_{PC}\matrixf{F}^{(1)}_{\bf k_1}.
\end{equation}
Here, $\matrixf{T}^{\rm red (2)}_{PC}$ indicates that its dimension have been reduced by this operation. Diagonalizing  $\matrixf{T}^{\rm red (2)}_{PC}$ in this subspace yields a set of eigenvectors $\matrixf{F}^{(2)}_{\bf (k_1,k_2)}$. Finally, we obtain $\matrixf{T}^{(3)}_{PC}$ for the third PC lattice vector
\begin{equation}
\matrixf{T}^{\rm red (3)}_{PC} = \matrixf{F}^{\dagger(2)}_{\bf (k_1,k_2)}\matrixf{T}^{ (3)}_{PC}\matrixf{F}^{(2)}_{\bf (k_1,k_2)}
\end{equation}
and the $\matrixf{F}^{(3)}_{\bf (k_1,k_2,k_3)}$ in the reduced subspace. The full eigenvectors $\matrixf{F}_{\bf k}$ can then be constructed via:
\begin{equation}
    \matrixf{F}_{\bf k} = \matrixf{F}^{(3)}_{\bf k}\matrixf{F}^{(2)}_{\bf k}\matrixf{F}^{(1)}_{\bf k} \ .
\end{equation}
This approach is also applicable for unfolding between a SC and a PC with different shapes,~i.e., when a non-diagonal transformation matrix $\matrixf{M}$ is used, as long as the mapping relation and the transformation matrix are provided. In this context, let us emphasize that 
the employed ``PC'' notation in our band unfolding algorithm does not need to correspond to the physical primitive cell.
Any cell that is smaller than the SC and that can be transformed commensurately to the SC can be used as ``PC'' depending on the reference band structure one would like to compare to~\cite{boykin2005_alloy_bzu,boykin2007_alloy_bzu}.
Fig.~\ref{fig:GaN_4096atoms}(a) shows the band structure of zinc blende GaN for the PC, the conventional cell, and a 64-atom cubic SC, respectively. In this case, the band structure obtained from the 64-atom SC can be unfolded onto either the PC or the conventional cell BZ.

\begin{figure}[t]
	\center{\includegraphics[width=1\linewidth]{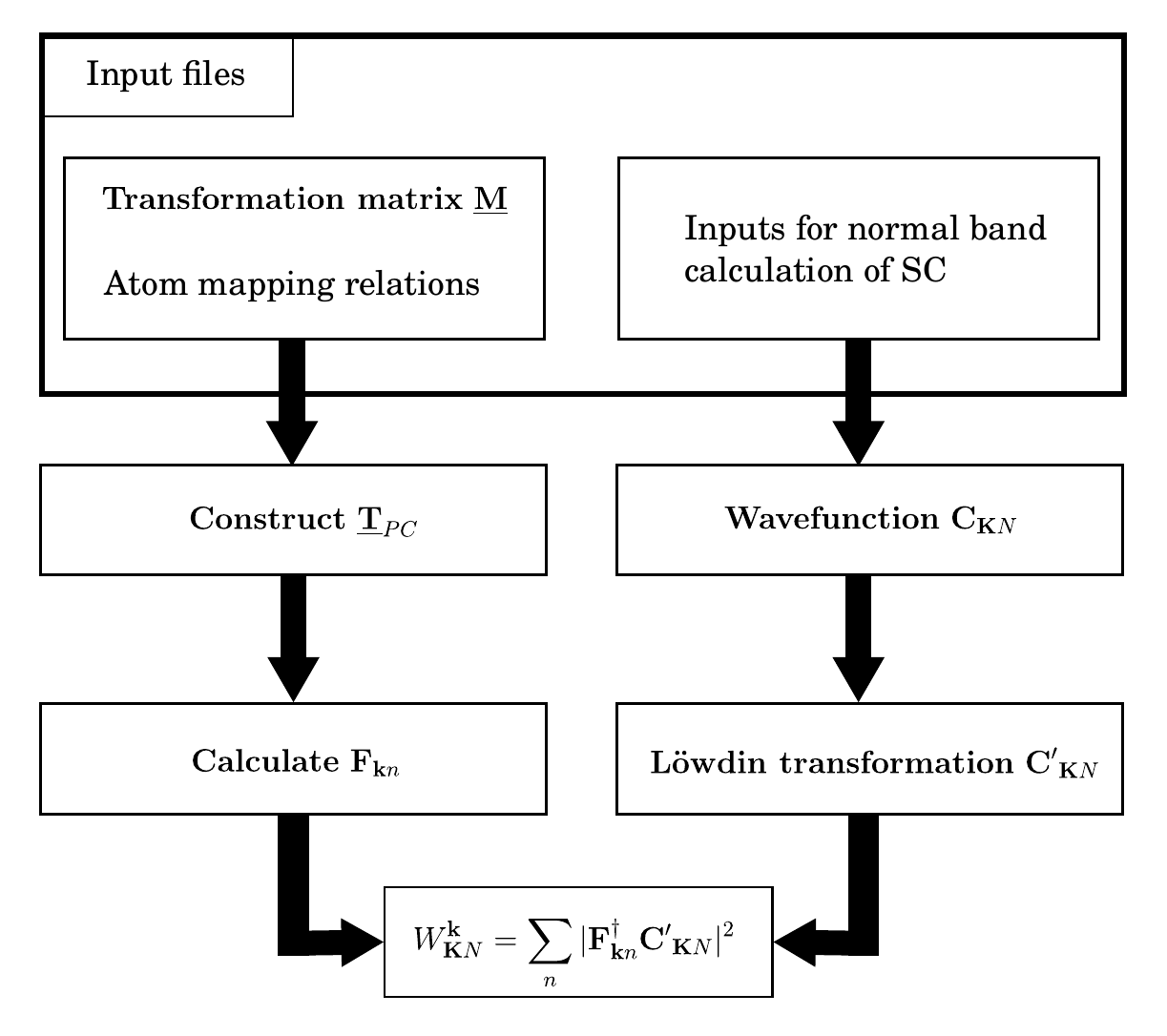}}
	\caption[]{Schematic code flow of the band unfolding implementation in {\tt FHI-aims}.
    }
	\label{fig:bzu_workflow}
\end{figure}

We implemented the band unfolding method described above as part of the native band-structure post-processing tool in {\tt FHI-aims}. The general workflow is illustrated in Fig.~\ref{fig:bzu_workflow}. In addition to the standard input tags required for SC band structure calculations, two additional pieces of information are necessary. These are the transformation matrix $\matrixf{M}$ between PC lattice vectors $\matrixf{a}$ and SC lattice vectors $\matrixf{A}$ and the mapping relation between atoms in the PC and the SC, which is used for constructing the translational operator~$\matrixf{T}$.
As in the case of a band structure calculation, the SC $\bf K$-points are independent in the band unfolding calculation, allowing for efficient parallelization. In our implementation in {\tt FHI-aims}, this is achieved by supporting both parallelization over {\bf k}-points using {\tt LAPACK} for the linear algebra and parallel linear algebra via {\tt ScaLAPACK}.
This enables efficient parallelization for varying system sizes and $\bf K$-point requirements and allows large supercells to be handled with minimal computational overhead.
For instance, Fig.~\ref{fig:GaN_4096atoms}(b) shows the unfolded band of a 4,096-atom GaN supercell with nearly 100,000 basis functions and 50,000 states. Our unfolding algorithm accounts for $\sim$ 45\% of the total runtime at the semi-local level of exchange-correlation with eight $\bf K$-points on the $\bf K$-path.

\begin{center}
    \begin{figure*}[ht!]
        \centering
        \includegraphics[width=1\linewidth]{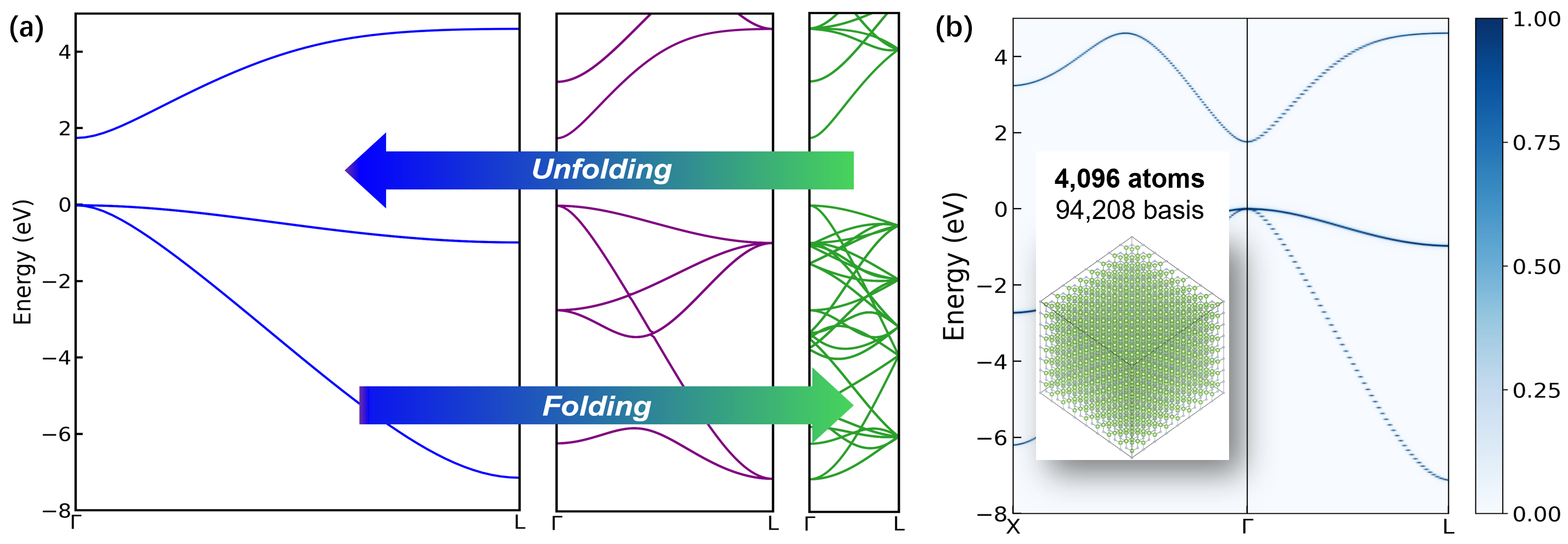}
        \caption{(a) Schematic illustration of band folding and unfolding in zinc blende GaN along the high-symmetry path $\Gamma$ {\textendash} L. Left: the primitive cell. Middle: the 8-atom conventional cell. Right: the 64-atom cubic supercell. (b) Unfolded band structure of a 4096-atom cubic GaN supercell along the high-symmetry path X {\textendash} $\Gamma$ {\textendash} L.}
        \label{fig:GaN_4096atoms}
    \end{figure*}
\end{center}

Note that the current implementation relies on the assumption that the number of basis functions in the supercell and primitive cell are commensurate. However, the presented derivation and methodology is general and equally applicable to systems with vacancies, interstitial atoms, alloys, or other systems where the number of basis functions in the supercell and primitive cell are not directly divisible. As discussed above in Sec.~\ref{sec: analytical T_PC}, this can be achieved by introducing placeholder orbitals when constructing the primitive cell translational operator. Similarly, our unfolding approach is also applicable to other (quasi-)particle energy spectrum, such as phonon or magnon dispersions.

\subsection{Temperature-Dependent Spectral Functions}
\label{sec: spectral_function}
To shed light on the physical insights that band unfolding enable, we here consider its role and interpretation in terms of many body perturbation theory. In this framework, the  Green's function of an energy eigenstate $\ket{\psi_{{\bf k}n}}$ of the PC Hamiltonian $\matrixf{h}$ exhibits a single pole:
\begin{equation}
    G_n({\bf k},E) = \frac{1}{E-\epsilon_{{\bf k}n}+i\eta} \ ,
\end{equation}
where $\eta \to 0^+$ is an infinitesimal parameter.
In a disturbed SC, the original PC symmetry is broken, which is equivalent to adding a perturbation $\matrixf{V}$ to the original Hamiltonian: $\matrixf{H} = \matrixf{h} + \matrixf{V}$ that breaks the original translational symmetry. In the spectral representation, the original single pole in the Green's function of $\ket{\psi_{{\bf k}n}}$ splits into a plethora of poles~\cite{coleman_manybody}:
\begin{equation}
\label{eq: lehmann}
    G_n({\bf k},E) = \sum_{{\bf K}N}\frac{\braket{ \psi_{{\bf k}n} | \Psi_{{\bf K}N}} \braket{ \Psi_{{\bf K}N} | \psi_{{\bf k}n} } }{E-E_{{\bf K}N}+i\eta} \ .
\end{equation}
Here, $\ket{ \Psi_{{\bf K}N} }$ are the eigenstates of the disturbed SC Hamiltonian. The total pole strength remains unchanged, i.e.:
\begin{equation}
    \sum_{{\bf K}N} \braket{ \psi_{{\bf k}n} | \Psi_{{\bf K}N}} \braket{ \Psi_{{\bf K}N} | \psi_{{\bf k}n} } = 1 \ .
\end{equation}
Eq.~(\ref{eq: lehmann}) is also called the Lehmann representation of the Green's function~\cite{coleman_manybody,bruus_flensberg}.
The spectral function is defined as the imaginary part of the Green's function in the limit $\eta \to 0^+$:
\begin{eqnarray}
    A({\bf k}, E) &=& \frac{1}{\pi}\sum_{n}{\rm Im} G_n({\bf k}, E) \nonumber \\
    &=& \sum_{n{\bf K}N}|\braket{ \psi_{{\bf k}n} | \Psi_{{\bf K}N}}|^2\delta(E-E_{{\bf K}N}) \nonumber \\
    &=& \sum_{{\bf K}N} W^{\bf k}_{{\bf K}N}  \delta (E - E_{{\bf K}N}) \ ,
\end{eqnarray}
which is precisely the definition given earlier in Eq.~(\ref{eq:spectral function}). The spectral function provides a measure of how much the electronic states are affected by the perturbation. For a system at finite temperature $T$, the electronic spectral function can be computed as a canonical ensemble average~\cite{Zacharias2020,claudia_T_spectral_func_prl,nery_spectralfunction_feynmandiagram}:
\begin{equation}\label{eq:averaged_spectral_function}
    \braket{A({\bf k}, E)}_T = \lim_{I\to\infty}\sum^{I}_{i} A^i({\bf k}, E) \ ,
\end{equation}
where $i$ enumerates the $i$-th sample in phase space at temperature $T$. The width of  $\braket{A({\bf k}, E)}_T$ reflects the scattering of electron states at finite temperature and is proportional to the reciprocal lifetime of the electron states. Meanwhile, the peak position corresponds to the renormalized energy level at temperature $T$. 
A key advantage of this definition is that it does not rely on
perturbation theory, in which the spectral function is Lorentzian by construction~\cite{Giustino2017}. By contrast, in this approach we can take into account all orders of anharmonic effects and electron-vibrationial couplings as long as sufficient samples are chosen to correctly explore the potential energy surface. 
Consequently, $\braket{A({\bf k}, E)}_T$ can exhibit asymmetric, non-Lorentzian line-shape that captures strong anharmonic interactions, particularly at high temperatures.
This non-perturbative approach to calculate the spectral function had been proven to be equivalent to a Feynman expansion to all orders in perturbation theory~\cite{nery_spectralfunction_feynmandiagram}. 

In {\it ab initio} calculations, a finite number of samples $I$ need to be selected to effectively explore the phase space.
Ideally, these samples are obtained using path-integral {\it ab initio}~molecular dynamics\cite{tuckerman2023statistical,i-Pi}, which incorporates quantum-nuclear effects, or {\it ab initio}~molecular dynamics~(aiMD), which provides the correct statistics in the classical limit~\cite{Zacharias2020,quan2024carrier}.

\section{Application}
\label{sec: application}

To demonstrate the advantages of our implementation, we use it to compute vibrationally renormalized electronic spectral functions. The therein reflected temperature-dependent renormalization of the electronic structure and its band gap is an important and experimentally measurable manifestation of the coupling between vibrational and electronic degrees of freedom. For example, experiments have shown that the band gaps of many semiconductors, such as silicon~\cite{bludau1974_si_bgr,macfarlane1958_si_bgr}, diamond~\cite{logothetidis1992_diamond_bgr}, and $\alpha$-AlN~\cite{guo1994_aln_bgr,brunner1997_aln_bgr}, decrease with increasing temperature, a phenomenon known as Varshni effect~\cite{varshni_effect}. In some cases, however, the band gap can also increase with temperature~(inverse Varshni effect), e.g.,~in lead halide perovskites~\cite{perovskite_inverse_varshni, zacharias2023_npj_perovskite} and in copper halides~\cite{Copper_inverse_varshni}.
Even in the low-temperature limit, the experimentally observed band gaps do not approach the ones of the static systems with immobile atoms at their equilibrium positions -- a direct consequence of the zero-point vibrations of the nuclei~\cite{ponce2015_bgr_many_materials,Giustino2017}. These effects can be assessed from first principles through the calculation of temperature-dependent electronic spectral functions. Most commonly, such investigations employ many-body perturbation theory on top of the harmonic approximation to model the interactions between electrons and phonons,~e.g.,~the celebrated Allen-Heine-Cardona (AHC) formalism~\cite{Allen1976,Allen1981} and more advanced approaches that further generalized and extend the AHC approach~\cite{antonius2015_zpr_anharmonic,Giustino2017,zacharias2023_npj_perovskite,nery_spectralfunction_feynmandiagram}. To fully overcome the limits of perturbation theory,~i.e.,~to incorporate both anharmonic effects and couplings between electrons and vibrations to all orders, it is possible to apply band structure unfolding on top of aiMD simulations, as discussed in the context of Eq.~(\ref{eq:spectral function}) and as demonstrated for the highly-anharmonic perovskite SrTiO$_3$ over a wide range of temperatures~\cite{Zacharias2020}.
 
In this work, we use the latter approach to demonstrate the validity of the proposed band unfolding implementation. More specifically, we compute vibrationally renormalized electronic spectral functions at low temperature and at 300~K for CuI. This material is known for its large anharmonicity at room temperature, which is driven by
the spontaneous formation of short-lived Frenkel defects~\cite{florian_prl_2023,Knoop.2023}. 
As detailed below, this example calculations 
further substantiate the validity of the proposed unfolding technique for obtaining spectral functions both for the quasi-harmonic low-temperature perturbative and the strongly anharmonic regime, at which traditional perturbative approaches are inapplicable~\cite{Zacharias2020}.
It is worth noting that the practical calculation of non-perturbative temperature-dependent spectral functions involves converging several numerical and physical parameters, a discussion of which goes well beyond the scope of this work. The interested reader can find more details in the original publications of Zacharias {\it et al.}~\cite{Zacharias2020} as well as in Ref.~\cite{nikita_thesis}.

\begin{center}
    \begin{figure*}[ht!]
        \centering
        \includegraphics[width=1\linewidth]{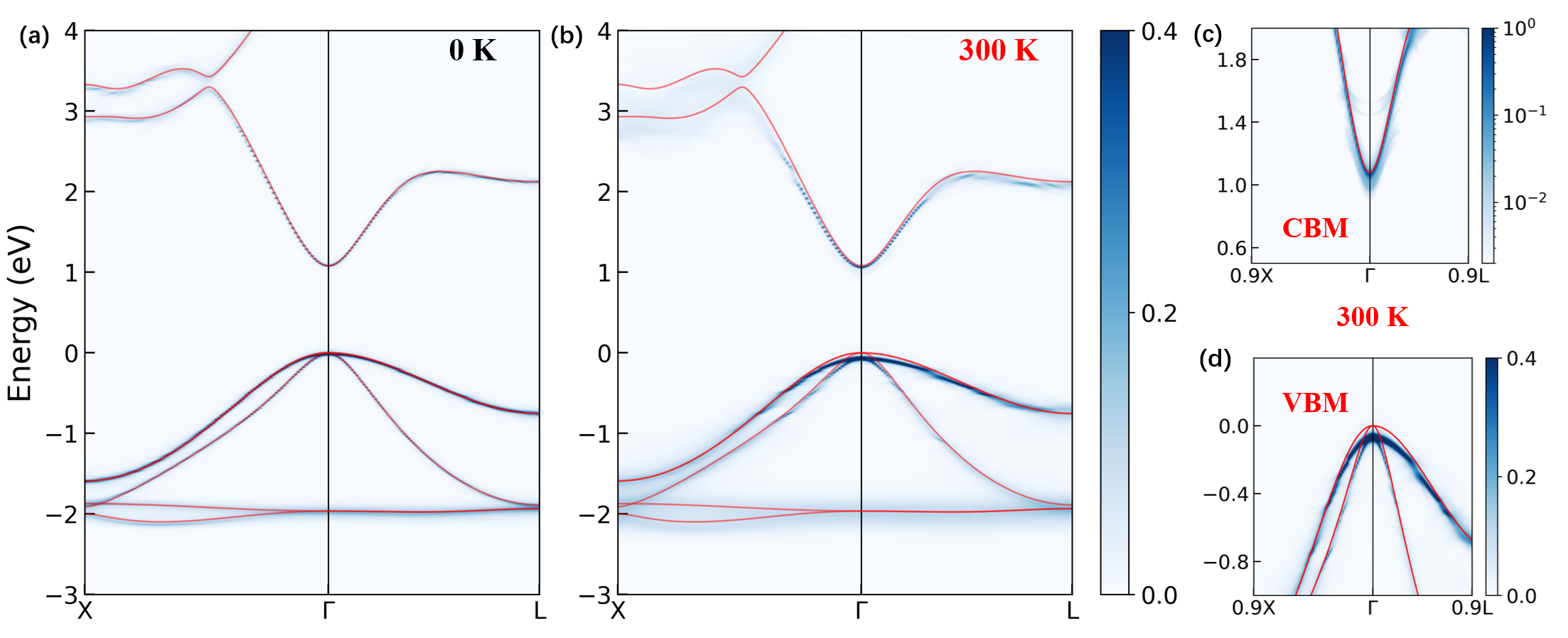}
        \caption{
        (a-b) Electronic spectral function of CuI at 0~K and at 300~K, see text for details. (c-d) Zoom-in of the spectral function near the conduction band minimum (CBM) and the valence band maximum (VBM). In all plots, the band structure of the static system,~i.e.,~without any vibrational effects, is shown in red. For consistency, the VBM of the static band structure is aligned to be zero in all panels.
        }
        \label{fig:arpes_CuI}
    \end{figure*}
\end{center}
All calculations were performed using the full-potential, all-electron NAO code {\tt FHI-aims}~\cite{Blum2009,aims_roadmap}. The PBEsol~\cite{Perdew2008} functional was used for describing exchange and correlation effects, ``\textit{light}'' defaults~\cite{Carbogno.2022} were chosen for the numerical settings and for the basis set, and a $15\times15\times15$ $\vec{k}$-point grid
was employed for Brillouin zone integrations in the primitive cell. The nuclear dynamics is modeled within a 216-atom cubic supercell 
both in aiMD and for the phonon calculations, which are performed using the finite difference approach~\cite{Parlinski1997} with  {\tt FHI-vibes}~\cite{Knoop2020_2} and its interface to {\tt phonopy}~\cite{Togo2015}. 
To quantify anharmonicity, we use the $\sigma^A$-metric proposed in Ref.~\cite{Knoop2020}, which
quantifies the fraction of interactions that stem from anharmonic effects on thermodynamic average as a function of temperature.
Accordingly, a $\sigma^A$ value around $\sim 0.5$, as found for CuI at 300~K, signals that 50\% of the
interactions in this systems originate from anharmonicity.
For the room-temperature case, the aiMD
trajectories available within the dataset~\footnote{\url{https://dx.doi.org/10.17172/NOMAD/2021.11.11-1}} were repurposed, while harmonic sampling~\cite{West2006} was used to generate atomic configurations in the low temperature regime. 
Lattice expansion at 300K is accounted for in the aiMD dataset in a fully anharmonic fashion, see Ref.~\cite{florian_thesis} for more details. 
This results in a $\sim 2 \%$ increase in cell-volume and, in turn, to a reduction of the band gap by $\sim 5 \%$ when neglecting the nuclear motion.
Also, all computational and physical settings used to generate the aiMD trajectory with {\tt FHI-aims} are the exact same ones used in our work for the band structure calculations.
The
harmonic sampling approach, which is only applicable when anharmonic effects are very small~\cite{Knoop2020,Zacharias2020}, has the advantage that quantum-nuclear effects can be straightforwardly incorporated.
In this 0 K reference calculations based on harmonic sampling, the lattice thermal expansion due to the zero-point vibration is neglected, since the respective volume expansion is anyhow very small~($\le 0.5$\%).
In our calculations, the energy eigenvalues for all configurations are output with respect to the internal zero ( i.e., the ${\bf G}=0$ component of the electrostatic potential ) of {\tt FHI-aims}, without alignment to the chemical potential. We then shift the energy reference to the valence band maximum of the CuI primitive cell, thereby placing the eigenvalues from different configurations on a consistent energy scale.
Eventually, spectral functions are obtained using 60 uncorrelated atomic configurations for each temperature via Eq.~(\ref{eq:spectral function}), whereby the therein appearing Dirac delta function is approximated by a Lorentzian with a broadness of 10~meV.
The chosen basis set, $\bf k$-grid and number of sampled, independent configurations ensure a convergence of the band-gap renormalization to $\pm 1$ meV. Similarly, the employed supercell size of 216-atoms ensures that the band-gap renormalization is converged up to a few meV.
Long-ranged, polar Fr{\"o}lich coupling is not captured by finite aiMD simulations, but can be included on top of them via correction terms~\cite{nery_frohlich,Zacharias2020}. By this means and by using the data published in Ref.~\cite{htp_frohlich}, we find that the Fr{\"o}lich coupling slightly decreases the band gap renormalization value by about 15~meV in CuI at 300~K, but would not change the observed trend.

Fig.~\ref{fig:arpes_CuI} shows the respective momentum-resolved spectral functions $\braket{A({\bf k}, E)}_T$ of zinc blende CuI. Here, panel~(a) features the spectral function obtained at $T=0$~K using a harmonic sampling with a quantum-statistics for the nuclei.
Strictly speaking, harmonic sampling is only valid in the limit of vanishing anharmonicity, since it neglects all anharmonic effects. Still, it can be employed in the limit of small anharmonicity~$\sigma^A\lesssim0.2$ to obtain qualitatively correct trends, as discussed in Ref.~\cite{Knoop2020}. Following these arguments, we use harmonic sampling to shed light on the 0K behavior of CuI. At this temperature, CuI exhibits moderate anharmonicity~($\sigma^A \approx 0.20$),
given the low anharmonicity metric~$\sigma^A \approx 0.20$ of CuI in this regime~\cite{Knoop2020}, which implies that only {20}\% of the interactions stem from anharmonic effects. In line with this, also the spectral function closely resembles the band structure obtained for the static system with immobile  atoms at equilibrium positions. Still, a minor, but finite shift and broadening of the electronic states due to zero-point atomic motion can be observed, as it is commonly the case in many-body perturbation theory calculations. In this low-temperature limit, the electron–vibrational coupling is weak, resulting only in small shifts that lead to a band-gap increase of {$10$~meV}. Similarly, only sharp electronic states with narrow linewidths are observed, which reflects that these states exhibit very long lifetimes. 

Conversely, the spectral function computed at 300~K and shown in Fig.~\ref{fig:arpes_CuI}(b) exhibits a drastically different behaviour with much more prominent shifts of the electronic states compared to the band structure of the static system, resulting in a much larger band-gap increase of {$56$~meV}.
{
Such an increase of the band gap, often denoted as the inverse Varshni effect, has also been observed for other copper halides~\cite{Copper_inverse_varshni}.
} Furthermore, a considerably stronger broadening is observed, indicating a much shorter lifetime of the electronic states. Indeed, some bands are smeared out to such a large degree that they become barely visible. 
In addition, Fig.~\ref{fig:arpes_CuI}(c) and (d) provide zoomed-in views of the spectral function at the band edges. This reveals that the spectral function at the conduction band minimum~(CBM) exhibits an asymmetric shape on the energy axis featuring shoulders and satellite peaks that deviates from the simple and symmetric Lorentzian shape that is inherent to many-body perturbation theory. All these effects are characteristic for higher-order couplings between electronic and vibrational degrees of freedom. And indeed, CuI exhibits much larger anharmonicity of $\sigma^A > 0.5$ at 300~K, indicating that more than 50\% of the interactions stem from anharmonic effects on thermodynamic average. As discussed in detail in Ref.~\cite{florian_prl_2023,Knoop.2023}, this strong anharmonicity originates from this system's tendency to spontaneously form of short-lived Frenkel defects. In turn, this leads to a breakdown of the harmonic phonon picture and, as demonstrated here, to a strong coupling of electronic and anharmonic vibrational degrees of freedom. 
{We emphasize that the satellite peak at 1.5~eV near the CBM can be associated to the occurrence of strongly anharmonic effects,~i.e.~to the spontaneous, but short-lived formation of a Frenkel defect~\cite{florian_prl_2023}, given that it is observed exclusively in configurations containing such Frenkel defects.}

\section{Conclusion}
\label{sec: conclusion}

In this work, we developed an efficient and robust band structure unfolding approach tailored for non-orthogonal LCAO-type of basis sets, addressing the challenges posed by the atomic-associated nature of AOs and enabling band unfolding calculations in all-electron implementations. To achieve this, we first simplified the expression of the unfolding weights using a L{\"o}wdin transformation. We then derived an analytical expression for the eigenvectors of the PC translational operator in the SC basis, enabling an efficient and accurate wavefunction projections. We implemented this approach in the all electron, NAO basis {\it ab initio} material simulation package {\tt FHI-aims}. 

Furthermore, we have demonstrated the practical efficiency and applicability of our approach. For instance, we have demonstrated the scalability of our implementation by investigating a 4,096-atom GaN supercell with nearly 100,000 basis functions, cf.~Fig.~\ref{fig:GaN_4096atoms}. Furthermore, we have showcased the usefulness of the developed unfolding approach in computing non-perturbative electronic spectral functions, as demonstrated in Fig.~\ref{fig:arpes_CuI} for CuI, a strongly anharmonic material. This further demonstrates the importance of this approach, since it gives access to regimes for which traditional many-body perturbation-theory models for electron-phonon coupling become inapplicable.

In summary, our band unfolding approach provides a powerful tool for investigating the underlying physics of complex materials that require supercell band structure calculations within the LCAO-type of basis sets.
A tutorial on band unfolding calculations using {\tt FHI-aims} is available at: 
\url{https://gitlab.com/FHI-aims-club/tutorials/band-unfolding}
}.

\begin{acknowledgments}
J.Q. acknowledges support from Prof. Angel Rubio and the Max-Planck Graduate Center for Quantum materials (MPGC-QM).
J.Q. would like to thank Mariana Rossi, Min-Ye Zhang, Chongxiao Fan {and Zekun Lou} for fruitful discussions.
N.R. acknowledges funding from the International Max Planck Research School for Elementary Processes in Physical Chemistry. 
M.S. acknowledges support by his TEC1p Advanced Grant (the European Research Council Horizon 2020 research and innovation programme, grant Agreement No. 740233).
\end{acknowledgments}

\section*{data availability}

All the calculations produced in this work are available on the Novel Materials Discovery (NOMAD) repository under~\footnote{\url{https://doi.org/10.17172/NOMAD/2026.01.06-1}}.

\begin{appendix}

\section{The Projection Operators}
\label{append: projector}
In this section, we demonstrate the equivalence between the projection operators defined in Eq.~(\ref{eq: projector linalg}) and (\ref{eq: projector group theory}) by comparing their corresponding unfolding weight $W^{\bf k}_{{\bf K}N}$. According to Eq.~(\ref{eq: projector linalg}), the unfolding weight of a SC state $\ket{{\bf K}N}$ on $\bf k$ is given by:
\begin{eqnarray}
    W^{\bf k}_{{\bf K}N} 
    &=& \braket{{\bf K}N|\matrixf{P}_{\bf k}|{\bf K}N}\\
    \label{eq: weight from linalg}
    &=& \sum_l \braket{{\bf K}N|{\bf k}l}\braket{{\bf k}l|{\bf K}N} \ .
\end{eqnarray}
On the other hand, using the projection operator defined in Eq.~(\ref{eq: projector group theory}) the unfolding weight becomes:
\begin{eqnarray}
    W^{\bf k}_{{\bf K}N} 
    &=& \braket{{\bf K}N|\matrixf{P}_{\bf k}|{\bf K}N}\\
    &=& \frac{1}{m}\sum^m_i \braket{{\bf K}N|\hat{{\bf t}}_i|{\bf K}N} e^{-i{\bf k}\cdot{\bf l}_i} \ .
\end{eqnarray}
Recall that $\ket{{\bf k}l}$ are eigenvectors of the translational operator $\hat{\bf t}$, therefore the operator $\hat{{\bf t}}_i$ can be expressed as:
\begin{eqnarray}
    \hat{{\bf t}}_i = \sum_l \ket{{\bf k}l} e^{i{\bf k}\cdot{\bf l}_i} \bra{{\bf k}l} \ .
\end{eqnarray}
Substituting this expression into Eq.~(\ref{eq: weight from linalg}), we find:
\begin{eqnarray}
    &&\sum_l \braket{{\bf K}N|{\bf k}l}\braket{{\bf k}l|{\bf K}N} \nonumber \\
    &=& 
    \bra{{\bf K}N}\Big ( \sum_l \ket{{\bf k}l}e^{i{\bf k}\cdot{\bf l}_i}e^{-i{\bf k}\cdot{\bf l}_i}\bra{{\bf k}l} \Big )\ket{{\bf K}N} \nonumber \\
    &=& \bra{{\bf K}N}\hat{\bf t}_i\ket{{\bf K}N} e^{-i{\bf k}\cdot{\bf l}_i} \ .
\end{eqnarray}
This relation holds for each of the $m$ PC translational operators $\hat{\bf t}_i$. Averaging over all of them, we finally obtain:
\begin{eqnarray}
    W^{\bf k}_{{\bf K}N} &=& \sum_l \braket{{\bf K}N|{\bf k}l}\braket{{\bf k}l|{\bf K}N} \\
    &=& \frac{1}{m}\sum^m_i\sum_l \braket{{\bf K}N|{\bf k}l}\braket{{\bf k}l|{\bf K}N} \\
    &=& \frac{1}{m}\sum^m_i \braket{{\bf K}N|\hat{{\bf t}}_i|{\bf K}N} e^{-i{\bf k}\cdot{\bf l}_i} \ .
\end{eqnarray}
This confirms the equivalence of the two projection operator definitions.

\section{Matrix Representation in a Non-Orthogonal Basis}
\label{appendix_matrix_non_ortho}
{Representing matrix equations in a non-orthogonal basis is significantly more complex than in an orthogonal one. However, the underlying mathematics is often introduced casually in the literature, leading to misunderstandings and confusion. To address this, we provide a detailed explanation of our approach for handling vector representations, matrix operations, and eigenvalue problems in a non-orthogonal basis in this section.

In a general non-orthogonal basis $\{ \ket{\phi_i} \}$ with overlap $S_{ij} = \braket{{\phi_i}|{\phi_j}}$, a vector ${\bf v}$ can be expressed as:
$${\bf v} = \sum_i v_i \ket{{\phi_i}} = 
\begin{pmatrix}
    \ket{\phi_1} & \ket{\phi_2} & \cdots & \ket{\phi_n}
\end{pmatrix} 
\begin{pmatrix}
    v_1 \\
    v_2 \\
    \vdots \\
    v_n \\
\end{pmatrix}
\ ,$$
where $v_i$ are coordinates in this basis set.
We can define a matrix(operator) $\matrixf{M}$ that maps ${\bf v}$ onto another vector $\bf w$:
$${\bf w} = \sum_i w_i\ket{{\phi_i}} = 
\begin{pmatrix}
    \ket{\phi_1} & \ket{\phi_2} & \cdots & \ket{\phi_n}
\end{pmatrix} 
\begin{pmatrix}
    w_1 \\
    w_2 \\
    \vdots \\
    w_n \\
\end{pmatrix}
$$ 
as $\matrixf{M}{\bf v}={\bf w}$. 
The corresponding matrix elements $M_{ij}$ in this basis are defined as:
\begin{eqnarray} \label{m_ij_mapping}
    \matrixf{M} \ket{\phi_j} &=& \ket{\phi'_j}=
    \begin{pmatrix}
        \ket{\phi_1} & \ket{\phi_2} & \cdots & \ket{\phi_n}
    \end{pmatrix} 
    \begin{pmatrix}
        M_{1j} \\
        M_{2j} \\
        \vdots \\
        M_{nj} \\
    \end{pmatrix} \nonumber\\
    &=& 
    \sum_i \ket{\phi_i} \ M_{ij}  \ .
\end{eqnarray}
As a consequence, the expansion coefficients for $\bf w$ are:
\begin{equation}
    w_i = \sum_{j} M_{ij} v_j \ .
\end{equation}
Note that, the definition of the matrix elements $M_{ij}$ above does not require  orthogonality of the basis set. It is defined only by the abstract mapping relation between vectors in linear algebra.

However, the matrix element $M_{ij}$ in this case are not given by $\braket{{\phi_i}|\matrixf{M}|{\phi_j}}$, as they would be in an orthogonal basis! This can be easily seen from:
\begin{eqnarray} 
     \braket{{\phi_i}|\matrixf{M}|{\phi_j}} &=& \sum_{k} \bra{\phi_i}M_{kj}\ket{\phi_k}\\
    &=& \sum_{k} S_{ik}M_{kj} = (SM)_{ij} \ .
\end{eqnarray}
Thus, in a non-orthogonal basis, the matrix element $M_{ij}$ defined by the mapping relation (\ref{m_ij_mapping}) is not equivalent to the inner product $\braket{{\phi_i}|\matrixf{M}|{\phi_j}}$. Some literature indiscriminately use both $M_{ij}$ and $\braket{{\phi_i}|\matrixf{M}|{\phi_j}}$ as matrix elements of $\matrixf{M}$ in a non-orthogonal basis, which leads to confusion. For completeness, 
we note that if the basis $\{ \ket{\phi_i} \}$ is orthogonal ($\matrixf{S} = \matrixf{I}$), the familiar relation $\braket{{\phi_i}|\matrixf{M}|{\phi_j}} = M_{ij}$ is recovered.

This raises a natural question: which quantity, $M_{ij}$ or $\braket{{\phi_i}|\matrixf{M}|{\phi_j}}$, should be used in an eigenvalue problem? The eigenvalue problem is generally formulated as:
\begin{equation} \label{ev_problem}
    \matrixf{A}{\psi_n} = \lambda_n {\psi_n} \ ,
\end{equation}
where ${\psi_n}$ is the eigenvector of the operator $\matrixf{A}$ corresponding to the eigenvalue $\lambda_n$. 
{Note that, eigenvectors ${\psi_m}$ and ${\psi_n}$ corresponding to different eigenvalues are always orthogonal, regardless of the basis set. }
Now we expand the above expression in a non-orthogonal basis set $\{ \ket{\phi_j} \}$, i.e., $\psi_n = \sum_j v_{nj}\ket{\phi_j}$, as:
\begin{equation}
    \sum_j v_{nj} \matrixf{A}\ket{\phi_j} = \lambda_n \sum_j v_{nj} \ket{\phi_j}.
\end{equation}
Applying $\bra{\phi_i}$ on the left results in the matrix equation:
\begin{equation}
    \sum_j v_{nj} \tilde{A}_{ij} = \lambda_n \sum_j v_{nj} S_{ij}
\end{equation}
or equivalently in matrix form:
\begin{equation} \label{ev_problem_lcao}
    \matrixf{\tilde{A}} {{\bf v}_n} = \lambda_n \matrixf{S} {{\bf v}_n} 
\end{equation}
with $\tilde{A}_{ij} = \bra{\phi_i}\matrixf{A}\ket{\phi_j}$. Thus, in a non-orthogonal basis, the standard eigenvalue problem (\ref{ev_problem}) transforms into the form Eq.~(\ref{ev_problem_lcao}), which is often called a generalized eigenvalue problem. As a result, the matrix elements that enter the generalized eigenvalue problem are given by the 
inner product $\braket{{\phi_i}|\matrixf{A}|{\phi_j}} = (SA)_{ij}$, rather than $A_{ij}$ itself. Furthermore, the orthonormal condition for the coefficients vectors ${\bf v}_n$ in the generalized eigenvalue problem follows from the requirement that all eigenvectors $\psi_n$ corresponding to different eigenvalues must be orthogonal:
\begin{eqnarray} 
     \delta_{mn} &=& \braket{\psi_m|\psi_n}  = \sum_{ij} v^*_{mi}S_{ij}v_{jn} \\
    &=& {{\bf v}^\dagger_m}\matrixf{S}{{\bf v}_n}.
\end{eqnarray}

All derivations presented in this work adhere to the aforementioned convention. Alternative methods for handling non-orthogonal bases, like introducing a reciprocal basis set~\cite{matrix_non_orthogonal} that incorporates the overlap matrix to express expansion coefficients, will not be discussed further in this paper.

}

\end{appendix}

\clearpage

%

\end{document}